\documentclass[usenatbib]{mnras}

\usepackage{amsmath,amssymb,gensymb}
\usepackage{multicol,tabularx}
\usepackage{enumitem}
\usepackage{graphicx,xcolor}

\usepackage[T1]{fontenc}
\usepackage{ae,aecompl}
\usepackage{txfonts}
\usepackage{color}
\usepackage{dsfont}
\usepackage{booktabs,multirow,subcaption,threeparttable}
\usepackage[capitalize]{cleveref}

\newcommand{\subtext}[2]{\ensuremath{#1_{\text{#2}}}} 
\newcommand{\linel}[2]{#1\,$\lambda$#2} 
\newcommand{\ergs}{\ifmmode {\rm erg\,s}^{-1} \else erg\,s$^{-1}$\fi} 
\newcommand{\kms}{\ifmmode {\rm km\,s}^{-1} \else km\,s$^{-1}$\fi} 

\title[Determining Quasar Orientation]{Determining Quasar Orientation}
\author[S.~Yong et al.]{\parbox[t]{\textwidth}{\vspace{-1cm}
Suk Yee Yong$^{1,}$\thanks{E-mail: \texttt{syong1@student.unimelb.edu.au}},
Rachel L.~Webster$^{1}$,
Anthea L.~King$^{1}$,
Nicholas F.~Bate$^{1}$,
Kathleen Labrie$^{2}$,
and Matthew J.~O'Dowd$^{3,4,5}$}
\vspace{1ex} \\
$^{1}$School of Physics, University of Melbourne, Parkville, VIC 3010, Australia\\
$^{2}$Gemini Observatory, Hilo, HI 96720, USA\\
$^{3}$Department of Physics and Astronomy, Lehman College of the CUNY, Bronx, NY 10468, USA\\
$^{4}$Department of Astrophysics, American Museum of Natural History, Central Park West and 79th Street, NY 10024-5192, USA\\
$^{5}$The Graduate Center of the City University of New York, 365 Fifth Avenue, New York, NY 10016, USA
}%

\date{Accepted XXX. Received YYY; in original form ZZZ}
\pubyear{2019}

\begin{document}
\label{firstpage}
\pagerange{\pageref{firstpage}--\pageref{lastpage}}
\maketitle

\begin{abstract}
Since the discovery of active galactic nuclei (AGN) and their subclasses, a unification scheme of AGN has been long sought. Orientation-based unified models predict that some of the diversity within AGN subclasses can be explained by the different viewing angles of the observer. Several orientation categorisations have been suggested, but a widely applicable measure has yet to be found. Using the properties of the ultraviolet and optical broad emission lines of quasars, in particular the velocity offsets and line widths of high-ionisation \ion{C}{iv} and low-ionisation \ion{Mg}{ii} lines, a correlation has been measured. It is postulated that this correlation is due to the viewing angle of the observer. Comparison with other orientation tracers shows consistency with this interpretation. Using a simulation of a wide angle disk-wind model for the broad emission line region, we successfully replicate the observed correlation with inclination. Future more detailed modelling will not only enable improved accuracy in the determination of the viewing angle to individual AGN, but will also substantially increase our understanding of the emitting regions of AGN.
\end{abstract}


\maketitle

\section{Introduction} \label{sec:intro}

There has been a long-held belief that there should be a simple unifying physical model for quasars, similar to the primary characterisation of stars by their mass. However its form and shape has proven elusive. Quasars are seen as the unresolved cores of distant galaxies, powered by accreting supermassive black holes. It has been difficult to disentangle the geometry of the emitting regions as there seems to be a degeneracy between orientation and at least three, and possibly more, parameters, such as the black hole mass, accretion rate, embodied in the Eddington ratio, and luminosity \citep[e.g.,][]{Laor:2000}. The geometry of some of the inner components can depend on the listed parameters, for example, the possible dependence of the thickness of the accretion disk and the characteristics of the emitting region on the accretion rate and luminosity of the objects \citep[e.g.,][]{Collin+Hure:2001}. The similarities between individual quasars also suggests possible unification, providing the motivation to search for a simple physical model. The classic \citet{Urry+Padovani:1995} model of active galactic nuclei (AGN) is axisymmetric, powered by an accretion disk surrounding a black hole. Thus orientation is expected to affect the measurement of observables at all wavelengths.

Although radio-quiet AGN are made up of the majority of AGN population, compared to their counterpart the radio-loud AGN, orientation effects are harder to establish in these objects. Most orientation tracers have been applied to radio-loud sources, despite their effectiveness being debated \citep[e.g.,][]{VanGorkom+:2015}. The presence of a relativistic jet in radio-loud AGN enables the determination of their inclination angle based on the jet alignment and radio properties \citep[e.g.,][]{Orr+Browne:1982,Ghisellini+:1993,Wills+Brotherton:1995}. One such dichotomy for radio quasars is flat-/steep-spectrum division. Viewing close to pole-on along the jet axis provides a direct view of the relativistic jet associated with non-thermal emission. The object will appear as a BL Lacertae (BL Lac) object or flat-spectrum radio-loud quasar (FSRLQ), both of which are blazars. BL Lac objects are distinguished by their optically featureless continua with almost no emission lines. As the viewing angle moves away from the jet, the Doppler-boosted component of the jet is weaker and the object is seen as a steep-spectrum radio-loud quasar (SSRLQ). FSRLQs tend to be core-dominated with flat radio spectra, while SSRLQs are mainly lobe-dominated and exhibit steep radio spectra \citep{Orr+Browne:1982}.

A measure of the strength of a Doppler-boosted jet in radio-loud sources is the radio core dominance. As the angle between the jet and line-of-sight decreases (towards pole-on), the emission from the core increases \citep{Padovani+Urry:1992,Ghisellini+:1993}, and hence higher core dominance will be detected. This is also related to the radio jet morphology between core- and lobe-dominated radio-loud quasars. Several parametrisations of radio core dominance have been introduced. The most frequently used core dominance parameter is defined by the ratio between the core and lobe flux density, $R$ \citep{Orr+Browne:1982}. Assuming that the extended lobe emission is isotropic, the core flux density reflects the Doppler boosting of the beamed jet, and thus inducing orientation dependence. However, the scatter in the relation is found to be large, especially for low values of $R$, and different normalisations have been considered such as using the optical continuum flux density \citep{Wills+Brotherton:1995}, narrow line region luminosity \citep{Rawlings+Saunders:1991}, and 151\,MHz radio luminosity \citep{Willott+:1999}. Some authors prefer one over the other, but there is no consensus of the most effective normalisation \citep[see e.g.,][]{VanGorkom+:2015,Marin+Antonucci:2016}.

The observable characteristics of emission lines have also been shown to be associated with orientation. This is established in the context of Eigenvector 1 \citep[E1;][]{Boroson+Green:1992}, attributed to the strong anti-correlation between the strength of \ion{Fe}{ii} and \linel{[\ion{O}{iii}]}{5007}. Based on the E1 along with full width at half maximum (FWHM) of H$\beta$, a unification scheme defined by Eddington ratio and orientation has been developed \citep{Shen+Ho:2014}. Using 20\,000 quasars at low redshift $0.1 \leq z \leq 0.9$ from the Sloan Digital Sky Survey (SDSS), they mapped the quasar distribution in the E1 optical parameter space and deduced that hints of orientation are contained in width of H$\beta$ line at constant value of $R_{\ion{Fe}{ii}}$, given by the EW ratio of \ion{Fe}{ii} and H$\beta$ lines. At pole-on, the H$\beta$ FWHM is narrow and becomes broader as the inclination angle increases approaching edge-on \citep[e.g.,][]{Wills+Browne:1986,Boroson+Green:1992,Shen+Ho:2014}.

Another orientation dependence observable is the equivalent width (EW) of \linel{[\ion{O}{iii}]}{5007} line \citep{Risaliti+:2011}. They assumed that the line strength of this narrow emission line exhibits roughly isotropic emission, whereas the continuum emits anisotropically from the optically thick and geometrically thin accretion disk \citep{Shakura+Sunyaev:1973}. Hence, the flux of the optical continuum from the disk is expected to decrease with increasing angle from face-on to edge-on, while the variation in the line intensity is insignificant. By inspecting $\sim 6000$ optical SDSS quasar spectra, \citet{Risaliti+:2011} argued that the power-law tail slope of -3.5 of the EW distribution above 30\,\AA\ is a strong signature of orientation and indicates near edge-on quasar. Values below this may be due to the intrinsic variance due to factors such as the continuum properties and narrow line region structure. They also demonstrated that the EW of [\ion{O}{iii}] increases along with broader H$\beta$ width as the viewing angle moves towards the equatorial plane. A follow-up study with twice the sample size also supports the idea \citep{Bisogni+:2017}.

A subpopulation of quasars that display broad absorption lines (BALs) bluewards of the main ultraviolet (UV) broad emission lines \citep{Weymann+:1991}, such as \ion{C}{iv}, also seems to provide clues to orientation. BAL quasars are divided into three sub-groups depending on the ionisation transition in the absorption. High-ionisation BAL (HiBAL) quasars display absorption in the high-ionisation species, like \ion{C}{iv}, \ion{N}{v}, and \ion{Si}{iv}. Low-ionisation BAL (LoBAL) quasars are HiBALs with the addition of absorption from low-ionisation species, like \ion{Mg}{ii}, \ion{Al}{iii}, and \ion{Al}{ii}. Iron low-ionisation BAL (FeLoBAL) quasars are LoBALs with extra \ion{Fe}{ii} or \ion{Fe}{iii} absorption. These objects are rare \citep{Hewett+Foltz:2003,Reichard+:2003,Gibson+:2009}, with HiBALs, LoBALs, and FeLoBALs ordered in increasing rarity. BALs exhibit a broad absorption trough, which is often blueshifted with respect to the emission line, that hints to evidence of outflowing material.

One interpretation of the BAL phenomenon is in the context of orientation model, whereby BAL quasars are detected when the line-of-sight and outflow intersects a disk-wind within a narrow range of angle to account for the low fraction of BAL population \citep[e.g.,][]{Murray+:1995,Elvis:2004}. The constraint on the outflow angle is not yet well-determined. Findings from spectropolarimetric observations favour an equatorial outflow geometry \citep{Goodrich+Miller:1995,Cohen+:1995,Ogle+:1999}. This is challenged by the existence of radio-loud BAL quasars, including those with LoBALs, that possess a large brightness temperature \citep{Zhou+:2006,Ghosh+Punsly:2007}. From radio variability reasoning, they argued that the observer's sightline is closely aligned with the radio emission from the relativistic jet in the polar direction about $\sim 10\degree \text{--} 35\degree$. However, this trait also is manifested in radio-quiet quasars with weak radio emission, which could be explained by an optically thin free-free (bremsstrahlung) emission from the accretion disk wind model \citep{Blundell+Kuncic:2007}. Other evidence that prefers a polar BAL wind is the excess narrow H$\beta$ line width in LoBAL quasars since a large width would indicate an equatorial view using the assumption of H$\beta$ line width as orientation tracer \citep{Punsly+Zhang:2010}.

Despite extensive studies of many observable parameters, a clean measurement of the orientation of the accretion disk rotational axis to the line-of-sight has proven elusive. We describe a simple measurement that maps the angle-of-viewing to a quasar, unlocking one of the systematic variables to understanding the complexities of observed quasar properties. We propose an orientation indicator based on the velocity shifts and line width ratio of high-ionisation \ion{C}{iv} and low-ionisation \ion{Mg}{ii} lines. We interpret the correlation with inclination in the context of the quasar disk-wind model \citep{Yong+:2018}. The outline is as follows. \Cref{sec:dataset} describes the quasar sample employed. Our proposed quasar orientation indicator is presented in \cref{sec:diffvratiofwhmi}. Comparison with other orientation tracers and simulation are examined in \cref{sec:otheri} and \cref{sec:compsim}, respectively. Further discussion on our inclination mapping is in \cref{sec:robusti}. Lastly, \cref{sec:summary} provides the summary.

\section{Dataset} \label{sec:dataset}

We select quasar data set using the catalogue of quasar properties \citep{Shen+:2011} derived from the SDSS Data Release 7 Quasar \citep[DR7Q;][]{Schneider+:2010} catalogue. The data are publicly accessible online through the VizieR catalogue access tool\footnote{\url{http://vizier.u-strasbg.fr/viz-bin/VizieR}} \citep{Ochsenbein+:2000}.

The specifications on the sample and spectral line measurements for SDSS DR7Q \citep{Shen+:2011} catalogue are described in Sect.~2 and Sect.~3 of their paper. The broad emission lines are modelled using at least one Gaussian profile and are separated into their corresponding broad and narrow line components, except for \ion{C}{iv} which retains both components. The FWHMs are also computed in the process. The velocity shifts with respect to the systemic redshift for \ion{Mg}{ii} broad line component and \ion{C}{iv} line are estimated from the fitted model emission line peak. They evaluated the uncertainties in the spectral line measurements using a Monte Carlo approach. From the flux density errors, Gaussian noise is introduced to the original spectrum by generating 50 random mock spectra fitted by the same fitting procedure to yield distribution of individual spectral quantity. The measurement errors are calculated from the 68 per cent range at the median of the distribution.

We further limit our sample to spectra with \ion{C}{iv} and \ion{Mg}{ii} FWHM and velocity shift measurements available, and median signal-to-noise ratio (S/N) per pixel of $\geq 15$ in \ion{Mg}{ii} region at 2700--2900\,\AA\ and \ion{C}{iv} region at 1500--1600\,\AA. To minimise the induced scatter in the observed correlation, we further constrain our sample as follows. Due to the presence of broad absorption features in quasars with BALs, the FWHM measurements of these objects might be imprecise. Therefore, we limit our sample to be non-BAL quasars. Furthermore, we also remove anomalous values, which are plausibly outliers, using the interquartile range (IQR) method \citep{Tukey:1977} for outlier detection. The IQR represents the dispersion of a data set and is given by the difference between upper (third or 75\textsuperscript{th} percentile) and lower (first or 25\textsuperscript{th} percentile) quartiles, IQR=Q$_{3}-$Q$_{1}$. Measurements outside of 1.5 times the IQR are considered outliers and are omitted. The implications of excluding these points will be discussed later. This reduces our quasar sample to 1835, with redshifts $1.50 \leq z \leq 2.25$, excluding BAL quasars (228) and outliers (98).

\section{Velocity Shifts and Ratio Line Width as Orientation Indicator} \label{sec:diffvratiofwhmi}

Qualitatively, we expect that a number of the physical characteristics of line shape will result from differences in the angle-of-viewing to the quasar: line velocity-offsets, line widths and line asymmetries. In order to test these ideas, we concentrate on the emission lines that are observed in the optical, but are emitted at UV wavelengths. Two broad emission lines with little contamination from other emission lines close by, originating from the broad line region (BLR), are compared: the high-ionisation \ion{C}{iv} and low-ionisation \ion{Mg}{ii} lines. Although the \ion{Fe}{ii} multiplets can contaminate the \ion{Mg}{ii} line, the \ion{Fe}{ii} emissions do not significantly affect the peak or FWHM of the \ion{Mg}{ii} line. The \ion{C}{iv} and \ion{Mg}{ii} lines are emitted at rest wavelengths 1549\,\AA\ and 2798\,\AA\ respectively with ionisation potentials of 64.49\,eV and 15.04\,eV. Typically, the emission lines in quasars are broadened, with FWHM velocities ranging from $\sim 2000 \text{--} 10\,000\,\kms$. Physical constraints on the state of the emitting regions implies that these velocities must result from bulk motions of the emitting ions.

\begin{figure}
\centering
\includegraphics[width=0.48\textwidth]{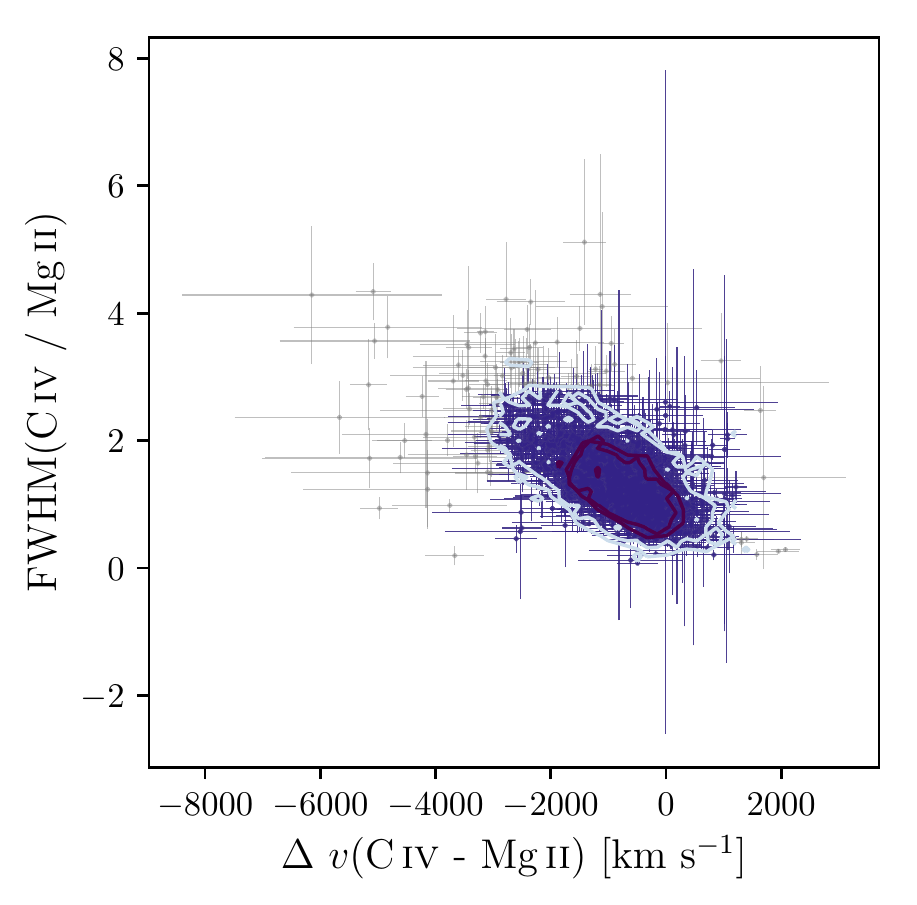}
\caption{Ratio of full width at half maximum, FWHM, against velocity shift, $\Delta v$, between \ion{C}{iv} and \ion{Mg}{ii}. We interpret the correlation to be attributed to inclination angle, with the upper left corner towards the lower right of the correlation indicating face-on to edge-on. The contours show the density of the sample at one- and two-sigma confidence levels. For comparison, the removed outliers are superimposed in small grey symbols.}
\label{fig:civmgiidiffv+ratiofwhm_dr7q}
\end{figure}

To examine trends in observable parameters, we use the quasar sample from SDSS DR7Q \citep{Shen+:2011} described in \cref{sec:dataset}. \Cref{fig:civmgiidiffv+ratiofwhm_dr7q} shows an anti-correlation between the velocity offset of the \ion{C}{iv} line with respect to the \ion{Mg}{ii} line, $\Delta v$(\ion{C}{iv}-\ion{Mg}{ii}), and the ratio of the FWHM of the two emission lines, FWHM(\ion{C}{iv}/\ion{Mg}{ii}). Based on a kinematical argument, we propose that the correlation is attributed to angle-of-viewing.

\begin{figure*}
\centering
\includegraphics[width=0.8\textwidth]{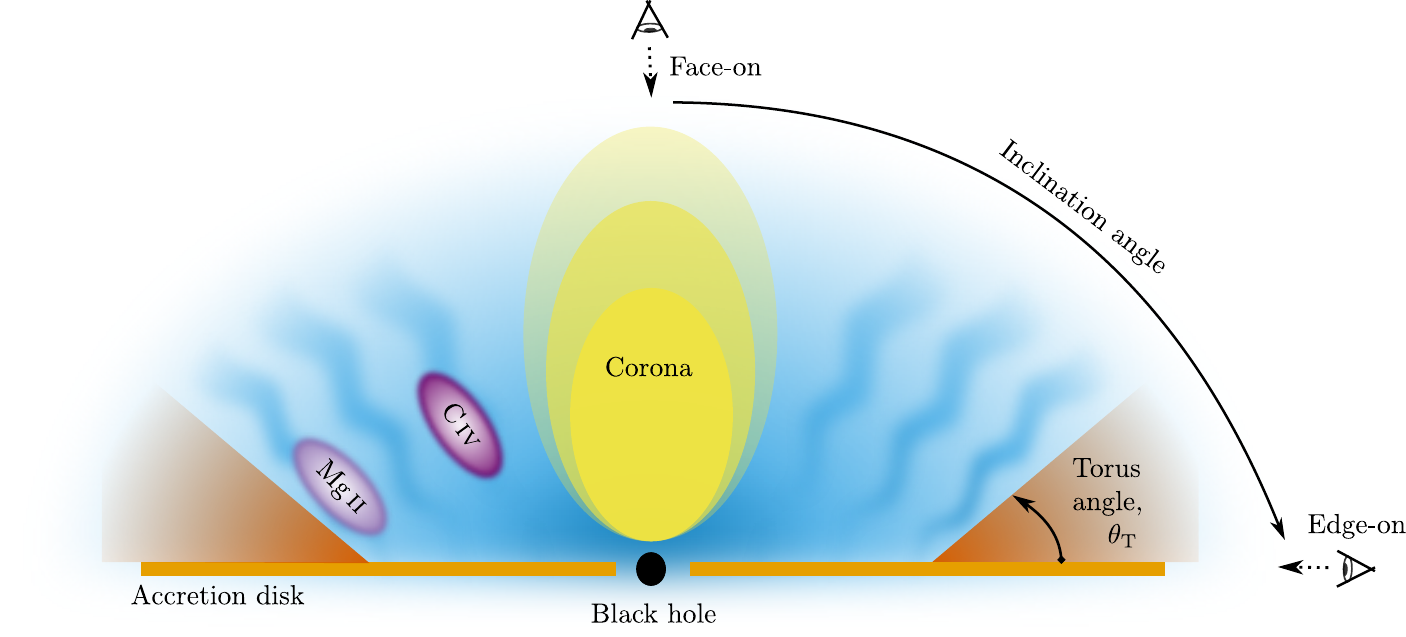}
\caption{Sketch of the key features of the disk-wind. The wind occupies a wide range of opening angles, with the high ionisation atoms at smaller inclinations and lower ionisation atoms closer to the putative torus.}
\label{fig:agnwindv3}
\end{figure*}

To measure the orientation of the quasar axis to the line-of-sight, a suitable model within which to interpret the key observables is required. In the UV/optical region of the spectrum, the broad emission lines are easily observable, with the broadening attributed to the velocity of the emitting ions projected onto the line-of-sight. To model the BLR, a kinematic disk-wind model is explored \citep{Yong+:2018}, based on previous quasar models \citep{Murray+:1995,Elvis:2004} but with some key differences. Importantly, a disk-wind covering a wide angle is required since there are similarities in emission line shapes between BAL and non-BAL quasars, which contradicts the wind arising from a narrow opening angle. \Cref{fig:agnwindv3} shows the features that differentiate our disk-wind model from others. In this model, ions leave the accretion disk at some radius, retaining the angular momentum associated with that radius (a rotational velocity), and then experience an outwards acceleration due to radiation pressure or a similar mechanism (a poloidal velocity). Thus, the trajectories of these ions will be outward spiralling paths, with different mixes of the two velocity components. Note that the effect of likely magnetic fields is ignored in this simple model. The opacity of the accretion disk, combined with a putative dusty torus will obscure the far-side of the BLR. Thus, only the forward-facing hemisphere of the quasar would be observed. Continuum emission is observed either directly from the accretion disk or re-processed emission from an atmosphere. It is expected that in the region close to the rotational axis, the ions will be completely ionised and so will not contribute to the emission line flux. This will be called the ionisation cone or corona. As the angle from the rotational axis increases, the density of the atmosphere or gas above the disk will increase, providing regions where the dominant emission will be from ions of decreasing ionisation. This is also a direct consequence of BLR stratification with different ionisation potential. Since the wind is likely to be launched in the proximity of the central ionising source, we expect high-ionisation lines such as the \ion{C}{iv} line to have a significant poloidal component, while retaining the velocity component from its rotational footprint. As the angle from the axis approaches the disk, emission from low-ionisation ions such as \ion{Mg}{ii} and H$\beta$ will be observed, largely reflecting the rotational velocity of their footprint. These outflowing components are termed the disk-wind, but the geometric location and kinematics of different ionic species will vary within the wind. Regions of higher density within the wind will be responsible for the absorption observed in BAL quasars. In particular, LoBALs should appear to be viewed from a preferential direction closer to the plane of the quasar. HiBAL quasars are still seen when the viewing angle intersects these streams, but they are not restricted to any preferential line-of-sight.

It is well-known that the \ion{C}{iv} line is mostly blueshifted with respect to other lines \citep[e.g.,][]{Gaskell:1982,Wilkes:1986,Espey+:1989,Tytler+Fan:1992,McIntosh+:1999,VandenBerk+:2001,Shen+:2016}. The \ion{Mg}{ii} emission is observed to be close to the systemic velocity of the quasar \citep{Hewett+Wild:2010}. Within the model described above, the \ion{Mg}{ii} line is expected to be emitted close to the disk and have only a small poloidal velocity component, while the \ion{C}{iv} line is emitted closer to rotational axis. Therefore, \ion{C}{iv} will reflect the projection of the poloidal or outflowing velocity along the line-of-sight, while \ion{Mg}{ii} will show predominantly the rotational velocity. The FWHM of each line will measure the projection of the bulk motion of the wind onto the line-of-sight. If these two ions are emitted in different parts of the wind, then the ratio of the FWHM of the two lines will reflect the different contributions to the two velocity components, projected along the line-of-sight. Thus, this ratio provides some constraints on the magnitude of the outflowing velocity, with respect to the rotational component at this particular angle of viewing.

The relation shown in \cref{fig:civmgiidiffv+ratiofwhm_dr7q} then reflects the strength of the poloidal and rotational velocity component, which can be used as a proxy to determine the inclination angle. Quasars viewed at face-on will be populating the top left-hand corner of the correlation and those views edge-on, towards the bottom right. Other physical parameters such as black hole mass and accretion rate, might determine the scatter in this relationship. Thus due to the intrinsic properties of the sources, the correlation might be expected to have a significant but natural dispersion. It is not a trivial matter to extract information on these intrinsic properties and is beyond the scope of the current paper.

\subsection{Methodology}

We conduct statistical analysis and linear regression techniques in order to investigate this relationship. Before fitting the correlation, we first rescale the data since the two variables have different magnitude which would result in a biased fit. This is done via z-score normalisation or standardisation, by centring the variable values at zero with a unit variance. In any case, rescaling data will not alter the general trend. Note that the original scales are used to produce the plots to illustrate the actual values.

Regression analysis is a valuable statistical tool for representing and examining the relationship between two or more variables. A simple linear regression model is when a linear relationship is considered with one dependent variable, $x$, and another independent variable, $y$, which gives the standard function $y=mx+c$, where $m$ and $c$ is the slope and intercept respectively. In linear regression technique, models with measurement errors attempt to minimise the variance along the fitted line. One measurement error model is the orthogonal regression model or also known as total least squares \citep{Adcock:1878,Golub+vanLoan:1980}, which assumes uncertainties in the two-dimensional data and minimises the squared orthogonal distance. This method treats the two variables symmetrically rather than depending on one another. For this reason, we choose to fit a linear line using orthogonal regression to the FWHM(\ion{C}{iv}/\ion{Mg}{ii}) versus $\Delta v$(\ion{C}{iv}-\ion{Mg}{ii}) correlation. To account for the measurement errors and intrinsic scatter in both velocity shifts and FWHM ratios, we perform the linear regression using bivariate correlated errors and intrinsic scatter \citep[BCES;][]{Akritas+Bershady:1996} method based on the publicly available Python module \citep{Nemmen+:2012}\footnote{\url{https://github.com/rsnemmen/BCES}}.

To assess the reliability of the fitted line if the same observation is done repeatedly, a bootstrap resampling method of the correlation is conducted. Random pairs of correlation are sampled with replacement from the original data set to assemble a new sample of the same size. The new sample is fitted with a linear regression model yielding its corresponding slope and intercept, and the whole process is iterated 1000 times. The median of the bootstrapped slopes and intercepts are taken as the best fit line. The 99.7 per cent confidence interval at the three standard deviation level of the slope and intercept are also computed. To quantify the fit, we calculate the residuals, which is distance between the observed and predicted values, and the corresponding mean squared error, where the error in this case is the residuals.

\begin{figure*}
\centering
\includegraphics[width=0.7\textwidth]{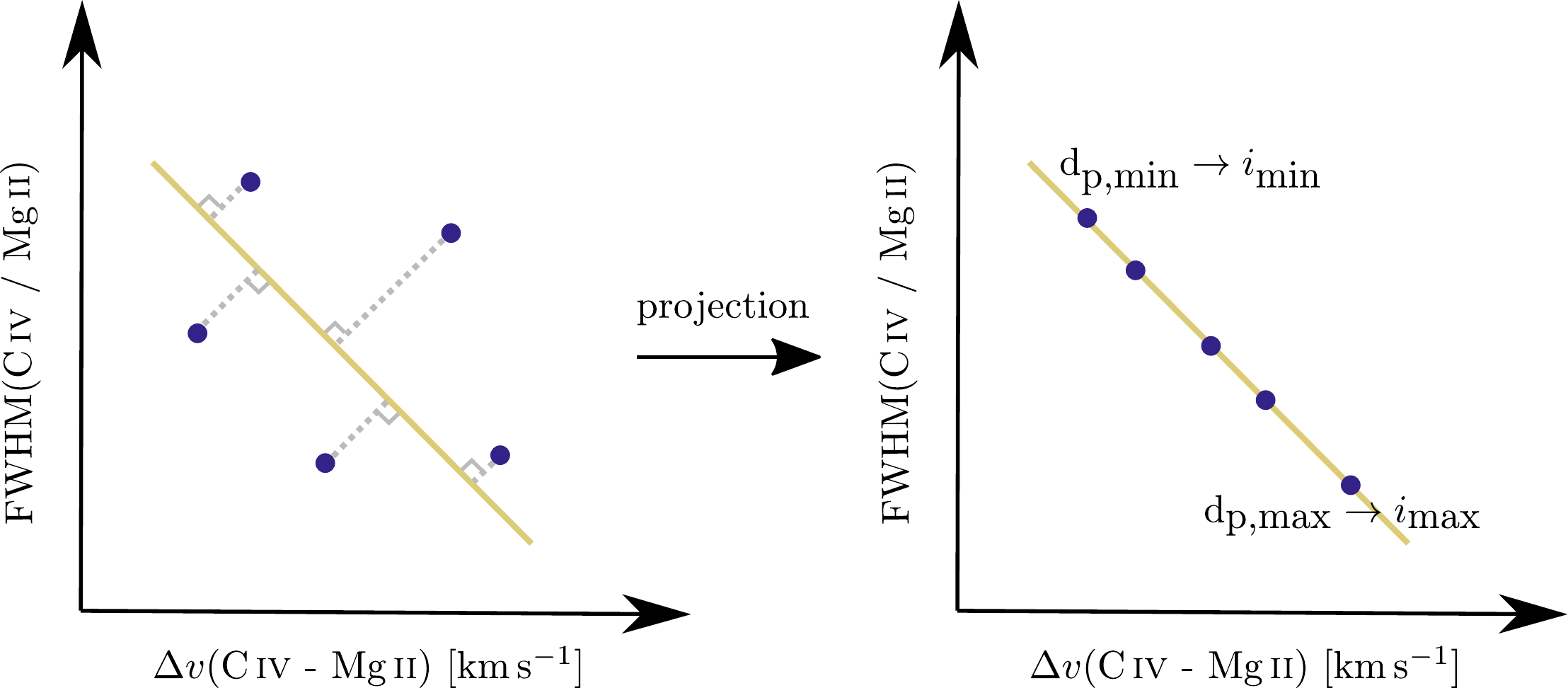}
\caption{Sketch of our proposed inclination mapping using the relationship between the velocity shift, $\Delta v$, and ratio full width at half maximum, FWHM, of \ion{C}{iv} and \ion{Mg}{ii}. The upper left corner towards the lower right of the correlation corresponds to face-on to edge-on viewing angle. {\em Left}: Each point is projected onto the best fit line as the projected distance, \subtext{d}{p}. The projection is performed in the perpendicular direction using orthogonal regression model, as shown by the dashed line. {\em Right}: The \subtext{d}{p} is then mapped to inclination angle, $i$, using \cref{eqn:mapdtoi}.}
\label{fig:dtoimap}
\end{figure*}

Determining the angle-of-viewing is clearly model dependent. We start by assuming a simplest possible mapping. For each quasar, the FWHM(\ion{C}{iv}/\ion{Mg}{ii}) and $\Delta v$(\ion{C}{iv}-\ion{Mg}{ii}) values are mapped to a projected distance along the best fit line. Based on the linear regression models, the mapping is performed in the perpendicular direction for orthogonal, vertical direction for $y$ on $x$, and horizontal direction for $x$ on $y$ regression. A distribution is generated from the projection. Assuming that the projected distance, \subtext{d}{p}, scales linearly with the inclination angle, $i$, the mapping is performed using
\begin{align}
i=90\degree \left[\frac{\subtext{d}{p}-\subtext{d}{p,min}}{\subtext{d}{p,max}-\subtext{d}{p,min}}\right],
\label{eqn:mapdtoi}
\end{align}
where \subtext{d}{p,min} and \subtext{d}{p,max} are the minimum and maximum limits of \subtext{d}{p}. A sketch of our proposed inclination angle mapping for orthogonal projection is portrayed in \cref{fig:dtoimap}. Indeed, the choice of projected scaling factor also contributes to the error in the inclination angle estimate. It is worth mentioning that the given prescription here is by no means univocal, but merely serves as a guide to whether it is viewed from polar, intermediate, or equatorial viewing angle. We further comment on the effects of our mapping procedure on the estimated viewing angle in \cref{ssec:errfits}.

The expected differential probability of observing a quasar at a particular inclination angle, $i$, defined from the rotation axis of the accretion disk, is given by $P(i)=\sin i$ with $0\degree \leq i \leq 90\degree$. The mapped inclination angle distribution is then normalised by the relative area on the sky to generate the `transparency distribution' as a function of inclination, i.e., the fraction of quasars whose inner regions are visible to the observer, given the orientation of the source and the presence of possible obscuring components, such as the torus.

\subsection{Results and Discussion}

\begin{figure}
\centering
\includegraphics[width=0.48\textwidth]{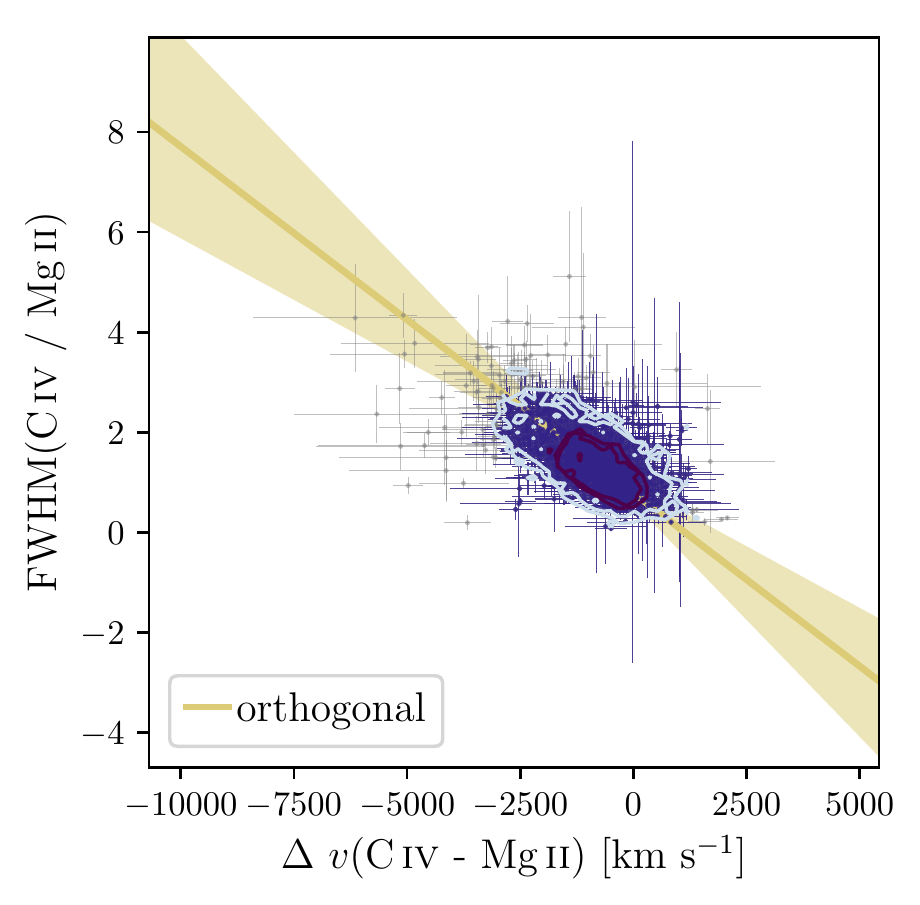}
\caption{Ratio of full width at half maximum, FWHM, against velocity shift, $\Delta v$, between \ion{C}{iv} and \ion{Mg}{ii} fitted using bivariate correlated errors and intrinsic scatter for orthogonal regression model. The fitted line is given in solid line with shaded region indicating the 99.7\% confidence interval. The contours show the density of the sample at one- and two-sigma confidence levels. For comparison, the removed outliers are superimposed in small grey symbols.}
\label{fig:civmgiidiffv+ratiofwhm_ort_dr7q}
\end{figure}

\begin{figure}
\centering
\includegraphics[width=0.48\textwidth]{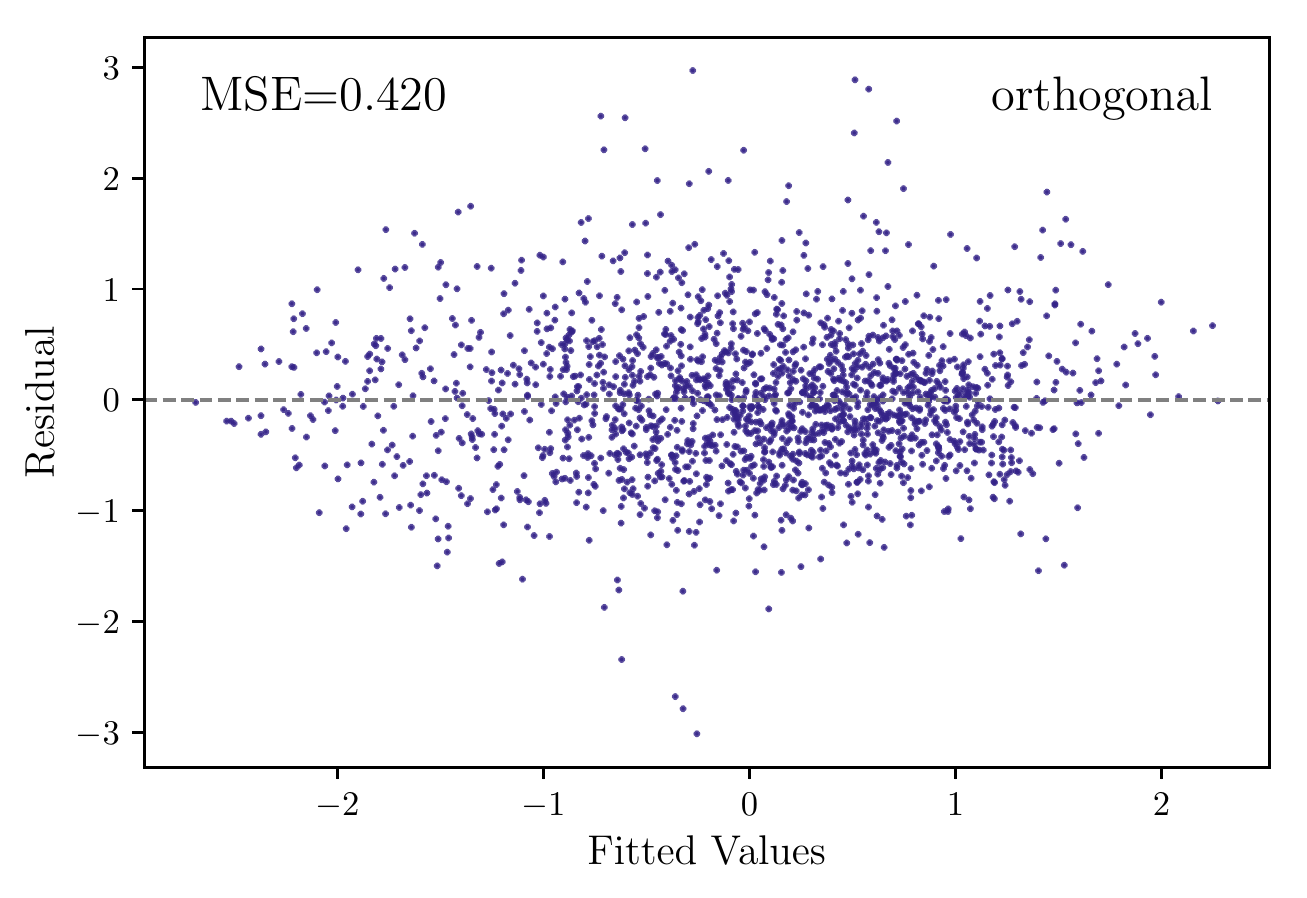}
\caption{Standardised residuals plot of the relation between ratio full width at half maximum and velocity shift of \ion{C}{iv} and \ion{Mg}{ii} using orthogonal regression model. The mean squared error (MSE) is listed on the upper left. The fitted line is shown in \cref{fig:civmgiidiffv+ratiofwhm_ort_dr7q}.}
\label{fig:res_orthogonal_dr7q}
\end{figure}

The BCES linear regression model fitting onto the $\Delta v$(\ion{C}{iv}-\ion{Mg}{ii}) and FWHM(\ion{C}{iv}/\ion{Mg}{ii}) correlation are displayed in \cref{fig:civmgiidiffv+ratiofwhm_ort_dr7q} along with the residuals in \cref{fig:res_orthogonal_dr7q}. The samples from SDSS DR7Q catalogues before (grey) and after (blue) removing the outliers are also shown. The fitted line parameters and their corresponding errors at 99.7 per cent confidence interval are presented in \cref{tab:regressionfit}. For comparison, the regression models without accounting for measurement errors are also provided. The Spearman's correlation coefficient, $r_{S}$, \citep{Spearman:1904} between $\Delta v$(\ion{C}{iv}-\ion{Mg}{ii}) and FWHM(\ion{C}{iv}/\ion{Mg}{ii}) and its corresponding correlation probability, $p_{S}$, are statistically significant with $r_{S}=-0.580$ and $p_{S} \ll 0.001$.

\begin{figure}
\centering
\includegraphics[width=0.48\textwidth]{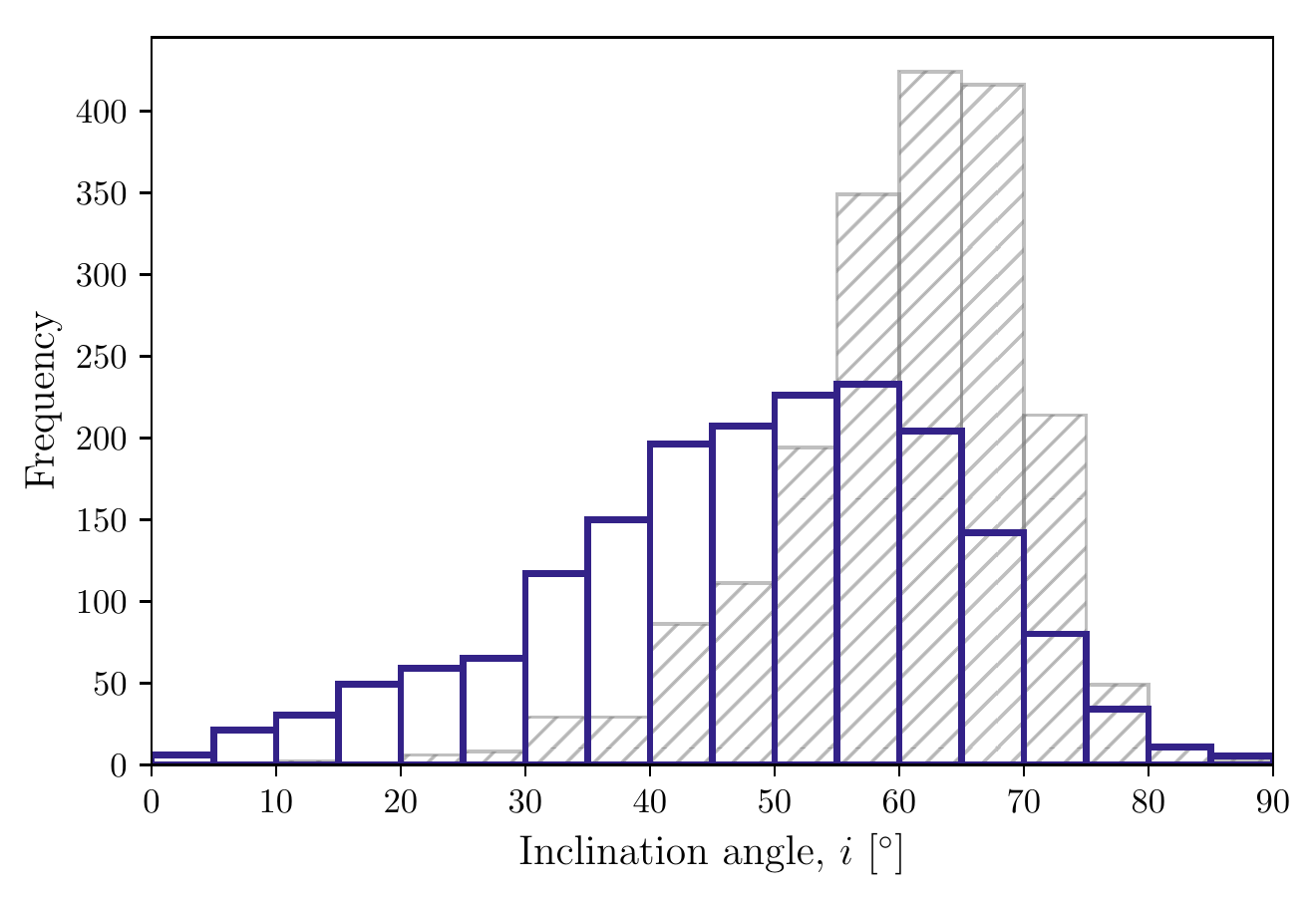}
\caption{Histograms of mapped velocity shift and ratio full width at half maximum of \ion{C}{iv} and \ion{Mg}{ii} plane onto inclination angle. The distribution of inclination angles is a linear projection of distance along the best fit line using orthogonal regression. The fitted line is shown in \cref{fig:civmgiidiffv+ratiofwhm_ort_dr7q}. For comparison, the distribution using samples including outliers is superimposed in hatch grey.}
\label{fig:histiproj_civmgiidiffv+ratiofwhm_ort_dr7q}
\end{figure}

\begin{figure}
\centering
\includegraphics[width=0.48\textwidth]{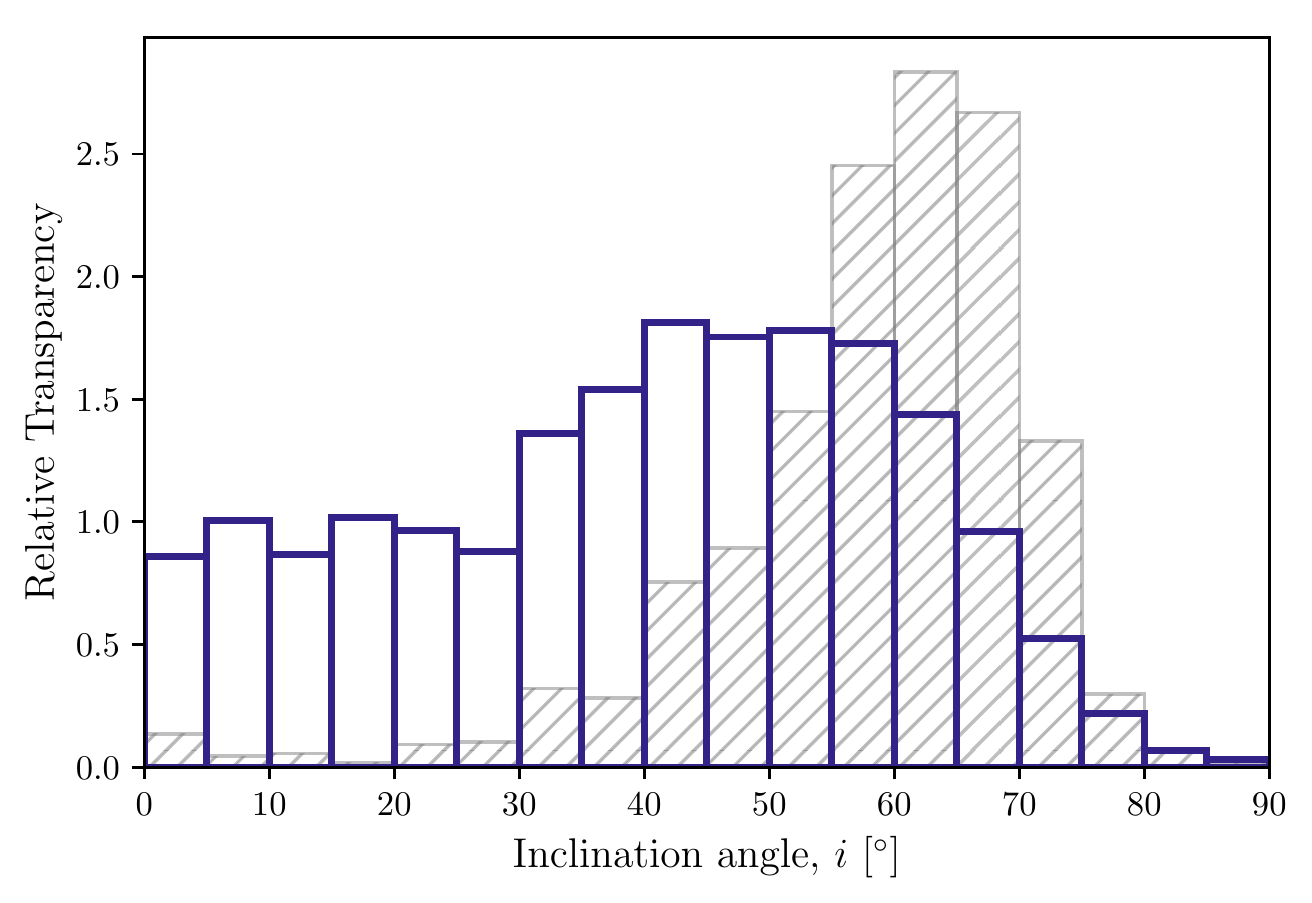}
\caption{Fraction of quasars along the line-of-sight normalised to area on the sky using velocity shift and ratio full width at half maximum of \ion{C}{iv} and \ion{Mg}{ii} mapping. The distributions of the projected distance mapped onto inclination are shown in \cref{fig:histiproj_civmgiidiffv+ratiofwhm_ort_dr7q}. For comparison, the distribution using samples including outliers is superimposed in hatch grey.}
\label{fig:transparencyi_civmgiidiffv+ratiofwhm_ort_dr7q}
\end{figure}

The distribution of our simple mapping from $\Delta v$(\ion{C}{iv}-\ion{Mg}{ii}) and FWHM(\ion{C}{iv}/\ion{Mg}{ii}) plane to inclination angle are demonstrated in \cref{fig:histiproj_civmgiidiffv+ratiofwhm_ort_dr7q}. The transparency plot that indicates the fraction of quasars that is not obstructing our line-of-sight is displayed in \cref{fig:transparencyi_civmgiidiffv+ratiofwhm_ort_dr7q}. The variation in the fitted slopes using different regression models are portrayed by the error bars.

The transparency is fairly constant at low inclination and peaks at intermediate angle, implying that most objects are seen at this range of direction. The scarcity of equatorial objects suggests an increase in obscuring material between the emitting region and the observer, which is possibly due to the presence of a torus. The dust distribution of the torus is supposedly clumpy, rather than smooth \citep[e.g.,][]{Krolik+Begelman:1988,Nenkova+:2002,Nenkova+:2008,Elitzur+Shlosman:2006}, such that emission from the ionising source is only partially blocked by the torus. This aspect is what distinguishes type 1 and 2 AGN, as have been established in the classic unification scheme of AGN \citep{Antonucci:1993,Urry+Padovani:1995}. Obscured AGN lack broad emission lines and are classified as type 2 AGN.

The boundary where there is a higher amount of obscuration along the line-of-sight is demonstrated by the drop-off in the distribution. This suggests that the angle of torus is $\subtext{\theta}{T} \sim 25\degree \text{--} 35\degree$ from the accretion disk plane using the samples without outliers (\cref{fig:transparencyi_civmgiidiffv+ratiofwhm_ort_dr7q}, thick blue). Other findings based on the fraction of two AGN types found the torus angle to be $\subtext{\theta}{T} \sim 30\degree \text{--} 40\degree$ \citep{Willott+:2000,Wilkes+:2013,Baldi+:2013,Marin+Antonucci:2016} and extend to $45\degree$ \citep{Barthel:1989}. These ranges of torus angle are consistent with our estimation.

However, the distribution of orientation angles obtained depends on the quality of the dataset analysed. This also affects the fitted regression, as illustrated by the underlying grey data points in \cref{fig:histiproj_civmgiidiffv+ratiofwhm_ort_dr7q} and \cref{fig:transparencyi_civmgiidiffv+ratiofwhm_ort_dr7q} when outliers are retained in the sample. In the SDSS DR7Q sample with outliers included, the underlying shape of both distributions are more negatively skewed compared to those when outliers are excluded. There are lower number of quasars at low inclination angles, and hence the transparency distribution is much flatter at low ends. The suggested torus angle is also smaller, about $\subtext{\theta}{T} \sim 20\degree$.

\begin{table*}
\centering
\caption{Velocity shift and ratio full width at half maximum of \ion{C}{iv} and \ion{Mg}{ii} correlation fitting using bivariate correlated errors and intrinsic scatter for orthogonal regression model. The linear fits of the expression FWHM(\ion{C}{iv}/\ion{Mg}{ii})$=m \Delta v$(\ion{C}{iv}-\ion{Mg}{ii})$+c$, where $m$ is the slope and $c$ is the intercept. The uncertainties in the fitted lines at 99.7\% confidence interval are computed using bootstrapping of 1000 samples.}
\label{tab:regressionfit}
\renewcommand{\arraystretch}{1.3} 
  \begin{threeparttable}
  \begin{tabular}{@{\extracolsep{4pt}}l c c c@{}}
    \toprule
    Catalogue & Slope & Intercept & Mean Squared Error \\
    \midrule
    S11 & $(-6.906 \substack{+1.838 \\ -1.901}) \times 10^{-4}$ & $0.798 \substack{+0.168 \\ -0.173}$ & 0.396 \\
    S11 [-O] & $(-6.918 \substack{+1.999 \\ -2.418}) \times 10^{-4}$ & $0.790 \substack{+0.165 \\ -0.200}$ & 0.420 \\
    \bottomrule
  \end{tabular}
  \begin{tablenotes}[flushleft]\footnotesize\setlength{\labelsep}{0pt}
    \item {\bfseries Note:} The quasar properties SDSS DR7Q catalogue \citep{Shen+:2011} as S11. Sample with outliers removed is indicated with [-O].
  \end{tablenotes}
  \end{threeparttable}
\end{table*}

\subsection{Error in Fits} \label{ssec:errfits}

Several aspects of the data can affect the mapping of inclination angle to the fitted line. Although we opt for high S/N quasar spectra, uncertainties in the spectral line measurements are inevitable. This can be due to systematic errors since the measurements from the catalogue are collected automatically, which might also be the cause of the redshifted velocity offsets seen.

Additionally, erroneous parameter values or outliers might be present as we did not manually check the accuracy of the automated measurements. This effectively results in a mapped inclination angle distribution with long tails, particularly on both ends at low and high inclinations. The data rescaling that we done, is an attempt to mitigate the contributions from the outliers.

Another key question that needs to be considered is the method for mapping from the data points to the regression line. For example, instead of using orthogonal regression model with projected points along the perpendicular direction, it could be projected along the vertical direction. The different gradients and projections produce different distributions of mapped projected distance to viewing angle.

As given by \cref{eqn:mapdtoi}, the normalisation for the histogram of mapped inclination is over $0\degree \text{--} 90\degree$. A different normalisation can be applied, for example $0\degree \text{--} 80\degree$ to account for the obscuration by the torus, although this would not significantly alter the distribution. The linear scaling between the projected distance and inclination might also be a simplification as the exact scaling is still unclear. Deducing a proper mapping will requires a thorough investigation and more insight on the physical properties and the relationship between them.

\section{Other Proposed Measurements of Inclination} \label{sec:otheri}

In the standard unified AGN scheme \citep{Urry+Padovani:1995}, the diversity of AGN classes is partly attributed to orientation. Various orientation indicators have been introduced in the literature, though most often accompanied by some limitations. We compare our proposed measurement of inclination with specific examples from the literature. We utilise the VizieR catalogue access tool to retrieve the data. Our choice of catalogues are limited to non-BAL objects within the SDSS coverage, unless mentioned otherwise, so that they can be cross-matched with the parent dataset from SDSS DR7Q properties catalogue \citep{Shen+:2011} that contain measurements of the spectral lines. This ensures a sufficient sample is acquired, though it does not guarantee that the spectra are of high S/N.

How can we compare the utility of our correlation to measure orientation with other indicators in the literature? We have not measured orientation, and so cannot make a direct comparison with metrics that measure a specific angle. However we can compare with correlations that claim to reflect orientation. We measure the Spearman's correlation coefficient, $r_{S}$, of our proposed correlation with other measures of inclination. The corresponding probability that two parameters are unrelated, $p_{S}$, is also measured, as listed in \cref{tab:spearman}. In all the sub-samples resulting from the cross-match with data in other bands, the FWHM(\ion{C}{iv}/\ion{Mg}{ii}) and $\Delta v(\ion{C}{iv}-\ion{Mg}{ii})$ correlation is significant at $p_{S}<0.05$.

\begin{table*}
\centering
\caption{Spearman's correlation coefficient for the investigated parameters.}
\label{tab:spearman}
  \begin{threeparttable}
  \begin{tabular}{@{\extracolsep{4pt}}l l c c c c@{}}
    \toprule
    Sample & Parameter & $\Delta v(\ion{C}{iv}-\ion{Mg}{ii})$ & $R$ & FWHM H$\beta$ & EW [\ion{O}{iii}] \\
    \midrule
    S11 [1933] & FWHM(\ion{C}{iv}/\ion{Mg}{ii}) & -0.608 &  &  &  \\
     &  & ($6.039 \times 10^{-196}$)\tnote{*} &  &  &  \\
    \midrule
    S11 [-O: 1835] & FWHM(\ion{C}{iv}/\ion{Mg}{ii}) & -0.580 &  &  &  \\
     &  & ($2.946 \times 10^{-165}$)\tnote{*} &  &  &  \\
    \midrule
    S11 [Cd+Ld: 231] & FWHM(\ion{C}{iv}/\ion{Mg}{ii}) & -0.454 &  &  &  \\
     &  & ($3.841 \times 10^{-13}$)\tnote{*} &  &  &  \\
    \midrule
    T06 x S11 [B: 72] & FWHM(\ion{C}{iv}/\ion{Mg}{ii}) & -0.264 &  &  &  \\
     &  & ($2.491 \times 10^{-2}$) &  &  &  \\
    \midrule
    K11 x S11 [1122] & FWHM(\ion{C}{iv}/\ion{Mg}{ii}) & -0.185 &  &  &  \\
     &  & ($4.217 \times 10^{-10}$)\tnote{*} &  &  &  \\
    \midrule
    K11 x S11 [L+T: 203] & FWHM(\ion{C}{iv}/\ion{Mg}{ii}) & -0.186 & 0.145 &  &  \\
     &  & ($7.813 \times 10^{-3}$) & ($3.870 \times 10^{-2}$) &  &  \\
     \cline{2-6}
     & $R$ & -0.222 &  &  &  \\
     &  & ($1.443 \times 10^{-3}$) &  &  &  \\
    \midrule
    T07 x S11 [47] & FWHM(\ion{C}{iv}/\ion{Mg}{ii}) & -0.425 &  &  &  \\
     &  & ($2.898 \times 10^{-3}$) &  &  &  \\
    \midrule
    T12 [70] & FWHM(\ion{C}{iv}/\ion{Mg}{ii}) & -0.271 &  & -0.711 & -0.411 \\
     &  & ($2.338 \times 10^{-2}$) &  & ($5.086 \times 10^{-12}$)\tnote{*} & ($4.036 \times 10^{-4}$)\tnote{*} \\
     \cline{2-6}
     & FWHM H$\beta$ & 0.203 &  &  & 0.534 \\
     &  & ($9.209 \times 10^{-2}$) &  &  & ($1.915 \times 10^{-6}$)\tnote{*} \\
     \cline{2-6}
     & EW [\ion{O}{iii}] & 0.194 &  & 0.534 &  \\
     &  & ($1.076 \times 10^{-1}$) &  & ($1.915 \times 10^{-6}$)\tnote{*} &  \\
    \midrule
    S12 x S16 [60] & FWHM(\ion{C}{iv}/\ion{Mg}{ii}) & -0.457 &  & -0.513 & -0.210 \\
     &  & ($2.418 \times 10^{-4}$)\tnote{*} &  & ($2.722 \times 10^{-5}$)\tnote{*} & ($1.068 \times 10^{-1}$) \\
     \cline{2-6}
     & FWHM H$\beta$ & 0.320 &  &  & 0.333 \\
     &  & ($1.263 \times 10^{-2}$) &  &  & ($9.300 \times 10^{-3}$) \\
     \cline{2-6}
     & EW [\ion{O}{iii}] & 0.091 &  & 0.333 &  \\
     &  & ($4.893 \times 10^{-1}$) &  & ($9.300 \times 10^{-3}$) &  \\
    \midrule
    T12, S12 x S16 [130] & FWHM(\ion{C}{iv}/\ion{Mg}{ii}) & -0.460 &  & -0.556 & -0.399 \\
     &  & ($3.780 \times 10^{-8}$)\tnote{*} &  & ($6.551 \times 10^{-12}$)\tnote{*} & ($2.637 \times 10^{-4}$)\tnote{*} \\
     \cline{2-6}
     & FWHM H$\beta$ & 0.196 &  &  & 0.375 \\
     &  & ($2.548 \times 10^{-2}$) &  &  & ($1.078 \times 10^{-5}$)\tnote{*} \\
     \cline{2-6}
     & EW [\ion{O}{iii}] & 0.304 &  & 0.375 &  \\
     &  & ($4.458 \times 10^{-4}$)\tnote{*} &  & ($1.078 \times 10^{-5}$)\tnote{*} &  \\
    \bottomrule
  \end{tabular}
  \begin{tablenotes}[flushleft]\footnotesize\setlength{\labelsep}{0pt}
    \item {\bfseries Note:} The catalogues are S11 for \citet{Shen+:2011}, T06 for \citet{Trump+:2006}, K11 for \citet{Kimball+:2011}, T07 for \citet{Turriziani+:2007}, T12 for \citet{Tang+:2012}, S12 for \citet{Shen+Liu:2012}, and S16 for \citet{Shen:2016}, with the character x indicates cross-match. The number of samples is in square brackets, with selected samples without outliers as -O, BAL as B, core- or lobe-dominated as Cd+Ld, and lobe or triple classes as L+T. The $p_{S}$ value is given in round brackets.
    \item[*] Highly significant at $p_{S}<0.001$.
  \end{tablenotes}
  \end{threeparttable}
\end{table*}

\subsection{Radio Morphology}

We use catalogues of quasar properties \citep{Shen+:2011}, radio properties of quasars \citep{Kimball+:2011}, and Radio Optical X-ray ASDC \citep[ROXA;][]{Turriziani+:2007}, that includes the radio morphology classification. The SDSS DR7Q properties catalogue \citep{Shen+:2011} contains radio properties obtained by matching the DR7Q catalogue to the Faint Images of the Radio Sky at Twenty centimetres \citep[FIRST;][]{White+:1997} catalogue using a 30'' matching radius. Radio quasars with single FIRST source within 30'' matching radius are rematched with higher matching radius of 5'' and are categorised into core-dominant radio quasars. Whereas, those with multiple FIRST sources within 5'' matching radius are lobe-dominant radio quasars. The rest are either undetected or outside FIRST area coverage.

The radio properties catalogue \citep{Kimball+:2011} has also utilised optical quasar spectra from SDSS and line measurements from SDSS DR7Q \citep{Shen+:2011} (see Sect.~2 of their paper). Their sub-sample of radio quasars are then classified via visual inspection using FIRST \citep{Becker+:1995} $2' \times 2'$ images and $4' \times 4'$ if the image has radio emission relative to the optical position of $\gtrsim 1'$. Depending on the location of the radio emission (see Sect.~3 of their paper), radio quasars are primarily sorted into core if emission is at optical position, jet if emission is from core and jet components, lobe if emission is from core and single lobe, and triple if emission from core and double lobe. The core and lobe 20\,cm flux densities of lobe and triple class are also provided. We then cross-match with the full sample from the quasar properties SDSS DR7Q catalogue \citep{Shen+:2011} to obtain the broad emission line FWHM and velocity shifts. We also select only those with clear identifiable morphology classification where two examiners concur with each other on the designation, as flagged in the catalogue. Additionally, we remove 3 points with extreme ratio FWHM of $>8$. The number of object in the sample is then 1122.

The ROXA catalogue \citep{Turriziani+:2007} includes identifications of blazars and FSRLQs. The selection and classification process are elaborated in Sect.~2 and Sect.~3 of their paper. Their sample is built based on multi-frequency approach using radio and X-ray surveys to attain the spectral slopes, through which possible candidates are identified. They then distinguished the radio sources by analysing the spectral energy distributions and optical spectra from the SDSS and 2dF surveys. We obtain 47 radio quasars with FSRLQ, BL Lac, or SSRLQ category after cross-matching their catalogue with SDSS DR7Q properties catalogue \citep{Shen+:2011}.

\begin{figure}
\includegraphics[width=0.48\textwidth,keepaspectratio]{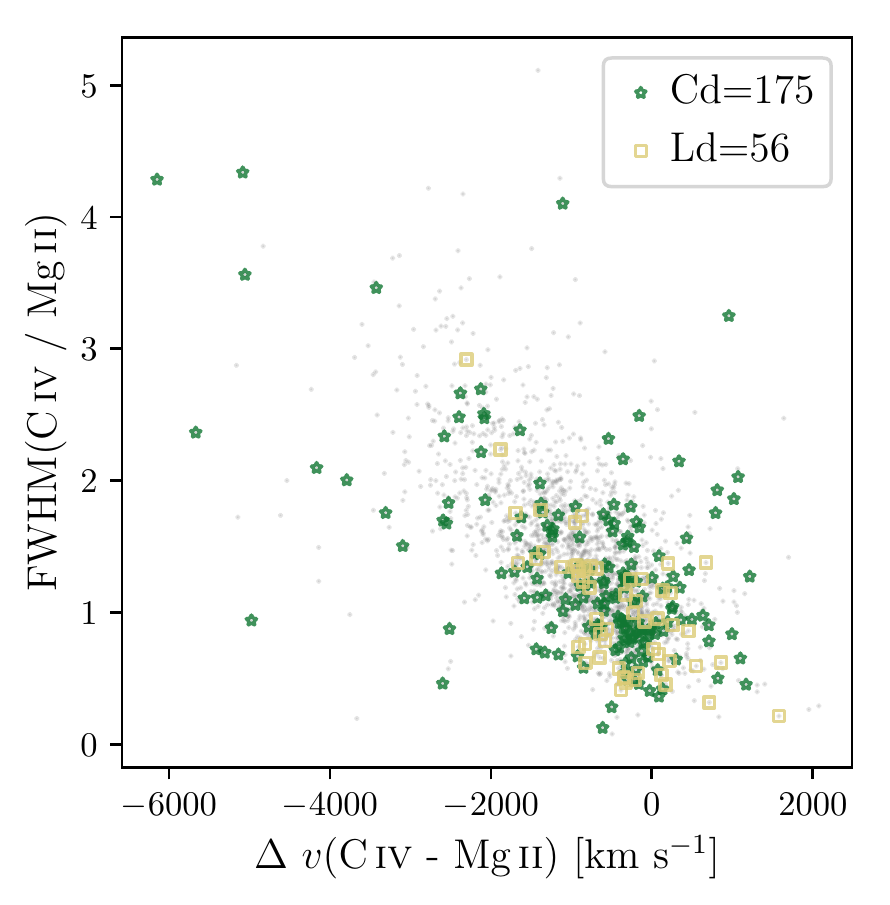}
\caption{Ratio of full width at half maximum, FWHM, against velocity shift, $\Delta v$, between \ion{C}{iv} and \ion{Mg}{ii}, with radio morphology. The quasar sample is obtained from \citet{Shen+:2011} SDSS DR7Q. Quasars classified as core-dominated (Cd) are indicated with green stars and lobe-dominated (Ld) with yellow squares. For comparison, their high spectral quality (high S/N) samples are superimposed in small grey symbols.}
\label{fig:civmgiidiffv+ratiofwhm+rm_s11xs11}
\end{figure}

\begin{figure}
\centering
\includegraphics[width=0.48\textwidth,keepaspectratio]{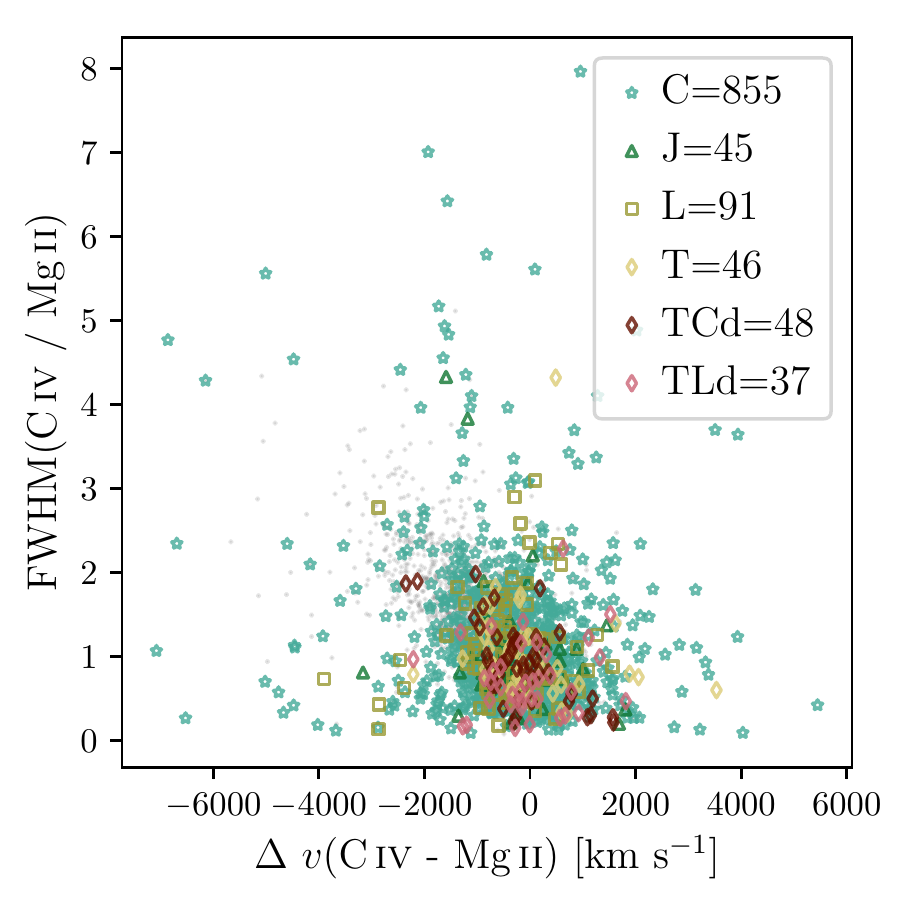}
\caption{Ratio of full width at half maximum, FWHM, against velocity shift, $\Delta v$, between \ion{C}{iv} and \ion{Mg}{ii}, with radio morphology. The radio morphology classifications are obtained from \citet{Kimball+:2011}. Quasars classified as core (C) are indicated with stars, jet (J) with triangles, lobe (L) with squares, and triple (T) with diamonds. Triple class quasars are subdivided into core-dominated triple (TCd) and lobe-dominated triple (TLd), denoted in different colours. The emission lines measurements are obtained from \citet{Shen+:2011} SDSS DR7Q catalogue. For comparison, their high spectral quality (high S/N) samples are superimposed in small grey symbols.}
\label{fig:civmgiidiffv+ratiofwhm+rm_k11xs11}
\end{figure}

\begin{figure}
\centering
\includegraphics[width=0.48\textwidth,keepaspectratio]{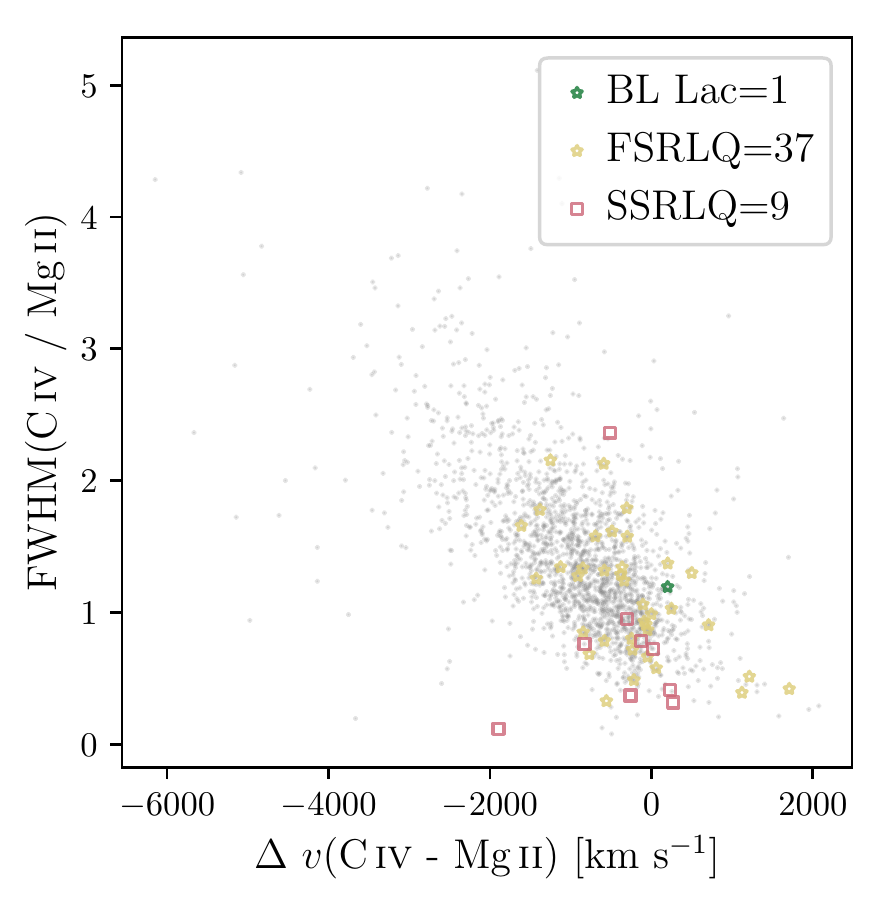}
\caption{Ratio of full width at half maximum, FWHM, against velocity shift, $\Delta v$, between \ion{C}{iv} and \ion{Mg}{ii}, with radio morphology. The radio morphology classifications are obtained from \citet{Turriziani+:2007}. Quasars classified as blazars are indicated with stars, including BL Lacertae (BL Lacs) in green and flat-spectrum radio-loud quasar (FSRLQ) in yellow, while steep-spectrum radio-loud quasars (SSRLQs) with pink squares. The emission lines measurements are obtained from \citet{Shen+:2011} SDSS DR7Q catalogue. For comparison, their high spectral quality (high S/N) samples are superimposed in small grey symbols.}
\label{fig:civmgiidiffv+ratiofwhm+rm_t07xs11}
\end{figure}

In \cref{fig:civmgiidiffv+ratiofwhm+rm_s11xs11,fig:civmgiidiffv+ratiofwhm+rm_k11xs11,fig:civmgiidiffv+ratiofwhm+rm_t07xs11}, we present the radio morphology overlaid onto the FWHM(\ion{C}{iv}/\ion{Mg}{ii}) versus $\Delta v$(\ion{C}{iv}-\ion{Mg}{ii}) distribution. Most data points are situated towards the lower right, which would signify intermediate to edge-on viewing angle in our simple inclination angle mapping. However, if we disregard the large dispersion in the relation at high FWHM(\ion{C}{iv}/\ion{Mg}{ii}) and blueshifts, and concentrate on the bulk of the dispersion, the difference between radio classes with inclination can be clearly identified. Core-dominated radio quasars are shifted towards the negative velocity shifts and higher FWHM ratio compared to those of lobe-dominated (\cref{fig:civmgiidiffv+ratiofwhm+rm_s11xs11,fig:civmgiidiffv+ratiofwhm+rm_k11xs11}). Similarly, most SSRLQs have the same trend as lobe-dominant radio quasars, while their counterpart FSRLQs as core-dominant (\cref{fig:civmgiidiffv+ratiofwhm+rm_t07xs11}). There is one BL Lac that seems to contradict with our assumption, though we are unable to statistically infer its true effect with only a data point. The jet and lobe classes in \cref{fig:civmgiidiffv+ratiofwhm+rm_k11xs11} are also clustered around the bottom right end of the trend.

Although the radio morphology classification provides a crude way to deduce the orientation of radio sources, generally they are in agreement with our simple approach of inferring the orientation. In our scheme, objects viewed along the line-of-sight approaching edge-on tend to have emission line properties that are less blueshifted with FWHM \ion{C}{iv} roughly equal or smaller than FWHM \ion{Mg}{ii}, which match with what we found for lobe-dominated SSRLQs. Meanwhile, core-dominated FSRLQs cover a wide range of FWHM ratios and velocity shifts, but mostly are blueshifted and at higher FWHM ratios.

\subsection{Radio Core Dominance}

We examine the trend in core dominance with our suggested model using the cross-matched sample between radio properties catalogue \citep{Kimball+:2011} and quasar properties SDSS DR7Q catalogue \citep{Shen+:2011} mentioned in the previous section. The core dominance in terms of ratio of core to lobe flux density at 20\,cm, $K$-corrected to its rest frame, is calculated for lobe and triple class radio quasars \citep{Kimball+:2011}:
\begin{align}
R=\frac{\subtext{S}{core}}{\subtext{S}{lobe}}(1+z)^{\subtext{\alpha}{lobe}-\subtext{\alpha}{core}},
\end{align}
where \subtext{S}{core} and \subtext{S}{lobe} are the observed 20\,cm core and lobe flux densities, and $\subtext{\alpha}{core}=0$ and $\subtext{\alpha}{lobe}=-0.8$ are the core and lobe spectral indexes.

\begin{figure}
\centering
\includegraphics[width=0.48\textwidth,keepaspectratio]{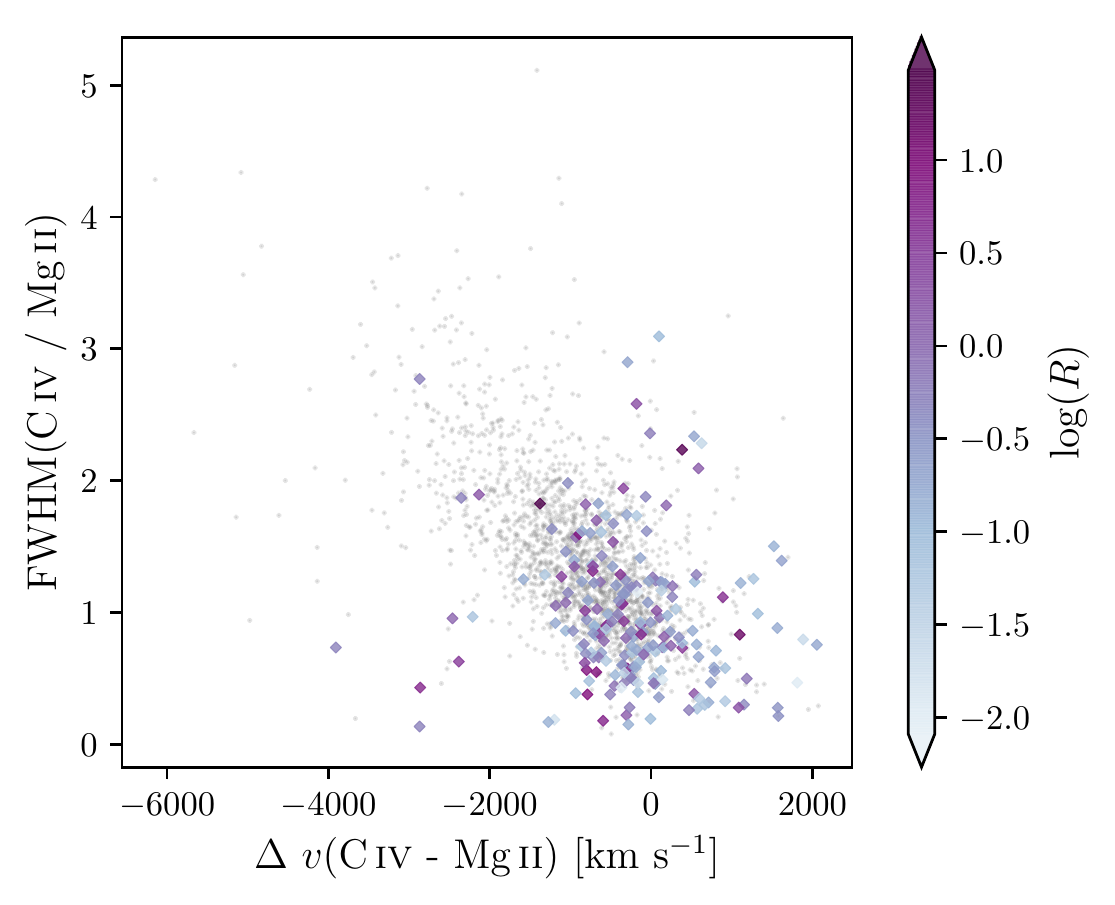}
\caption{Ratio of full width at half maximum, FWHM, against velocity shift, $\Delta v$, between \ion{C}{iv} and \ion{Mg}{ii}, with core dominance, $R$. The distribution is colour-coded by increasing colour gradient with $R$ using quasars sample of \citet{Kimball+:2011}. The emission lines measurements are obtained from \citet{Shen+:2011} SDSS DR7Q catalogue. For comparison, their high spectral quality (high S/N) samples are superimposed in small grey symbols.}
\label{fig:civmgiidiffv+ratiofwhm+rc_k11xs11}
\end{figure}

\Cref{fig:civmgiidiffv+ratiofwhm+rc_k11xs11} shows the core dominance in $R$ for the radio sources. The values of $R$ seem to be increasing, from the lower right to upper left of the ratio FWHM versus the velocity shift relation. This is also identified using the Spearman correlation coefficient (\cref{tab:spearman}). The parameter is significantly anti-correlated with $\Delta v$(\ion{C}{iv}-\ion{Mg}{ii}) at ($r_{S}=-0.222, p_{S}<0.01$) and correlated with FWHM(\ion{C}{iv}/\ion{Mg}{ii}) at ($r_{S}=0.145, p_{S}<0.05$). Although it has been argued that radio lobe flux density is not the best normalisation for radio core dominance \citep{Wills+Brotherton:1995}, our results are consistent with edge-on objects having lower $R$.

\subsection{Line Width of \texorpdfstring{H$\beta$}{Hb}}

We use two samples from the literature that provide emission line measurements covering the UV and optical regime. The first sample \citep[][hereafter T12]{Tang+:2012} contains bright quasars with redshift $z<1.4$, including radio-loud and radio-quiet, compiled from three subsamples, namely selected UV-excess Palomar-Green quasars, spectra with far-UV from Far Ultraviolet Spectroscopic Explorer and Hubble Space Telescope, and radio-loud quasars. Sect.~2 of their paper outlines the sample selection and spectral fitting procedure. They modelled the regions in the vicinity of \ion{C}{iv}, \ion{Mg}{ii}, and H$\beta$ with a power-law continuum. Two Gaussian components are applied to fit individual broad emission lines, except for H$\beta$ which requires an extra Gaussian to reflect the narrow line region emission. Using the fitted model, they calculated the properties of the emission lines. They also evaluated the velocity shifts between the fitted peak of the line and the systemic redshift. The spectral line measurements are listed in Table 3 and Table 4 of their paper. For our sample, we use 70 out of 85 objects that have complete emission line measurements for \ion{C}{iv}, \ion{Mg}{ii}, H$\beta$, and [\ion{O}{iii}].

The second sample consists of 60 luminous quasars at intermediate redshift of $z \sim 1.5\text{--}2.2$ that cover the \ion{C}{iv} to H$\beta$ region \citep{Shen+Liu:2012} and a subsequent follow-up survey that extends to cover [\ion{O}{iii}] region \citep{Shen:2016}. The selection of sample and spectral measurements are specified in Sect.~2 \citep{Shen:2016} and Sect.~3 \citep{Shen+Liu:2012} of their paper. They applied spectral fitting to the optical and near-infrared spectra to extract the continuum and line features. A pseudo-continuum is created by fitting a power-law continuum using UV and optical line templates, depending on the wavelength region. It is then subtracted from the original spectrum to yield the spectrum of emission lines. They fitted Gaussian profiles in logarithmic wavelength to model the broad and narrow components of the emission line. Using the fitted model, the emission line properties are derived, including the FWHMs of the broad line component. The relative velocity shifts are measured at the centroid of the narrow [\ion{O}{iii}], broad \ion{C}{iv}, \ion{Mg}{ii}, and H$\beta$. Since [\ion{O}{iii}] is not within the coverage in \citet{Shen+Liu:2012}, in our analysis, the H$\beta$ and [\ion{O}{iii}] line measurements are taken from \citet{Shen:2016}, while the \ion{C}{iv} and \ion{Mg}{ii} line properties are from \citet{Shen+Liu:2012} to be consistent. Hereafter, this cross-matched sample is the S12xS16 sample.

\begin{figure}
\centering
\includegraphics[width=0.48\textwidth,keepaspectratio]{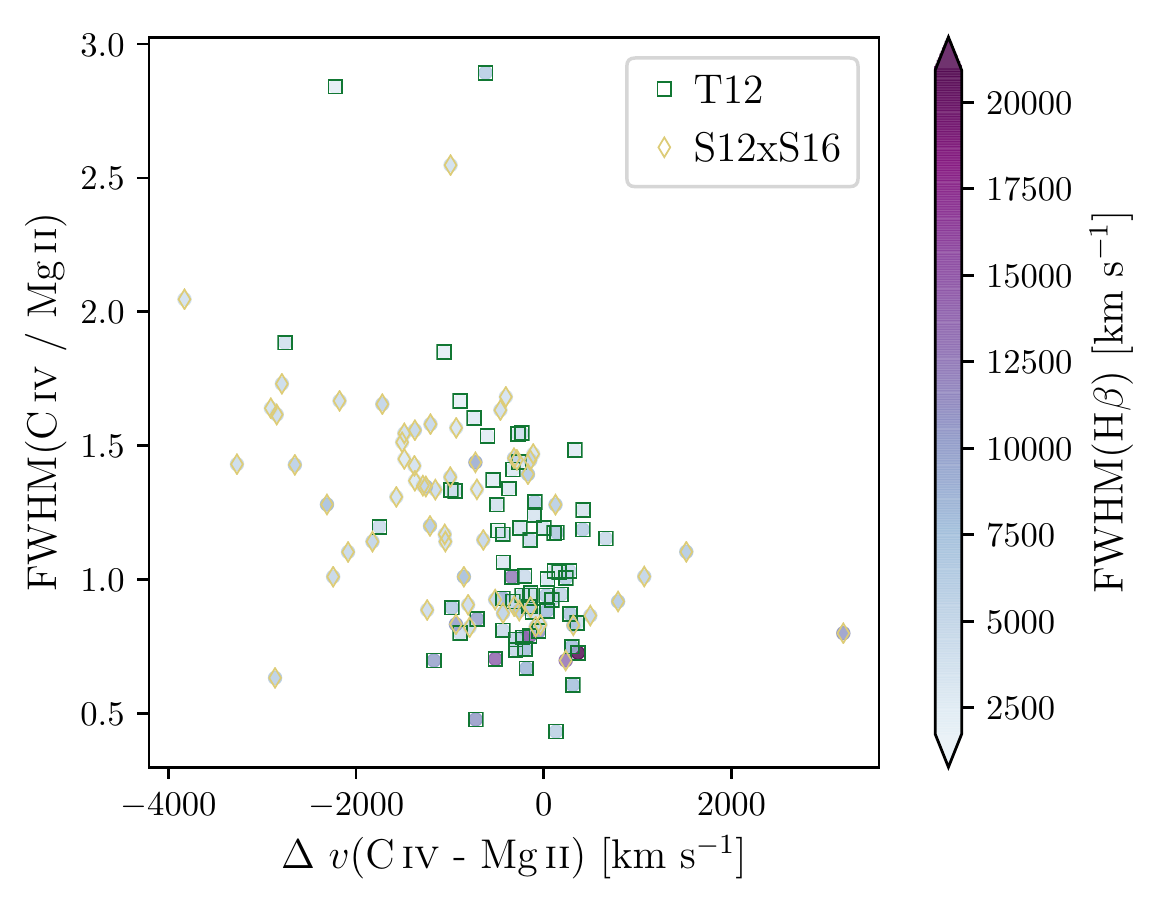}
\caption{Ratio of full width at half maximum, FWHM, against velocity shift, $\Delta v$, between \ion{C}{iv} and \ion{Mg}{ii}, with FWHM H$\beta$. The distribution is colour-coded by increasing colour gradient with FWHM H$\beta$ using quasars sample of T12 \citep{Tang+:2012} in green squares and S12xS16 \citep{Shen+Liu:2012,Shen:2016} in yellow diamonds.}
\label{fig:civmgiidiffv+ratiofwhm+hbfwhm_t12s1216}
\end{figure}

The distribution of H$\beta$ FWHM onto the $\Delta v$(\ion{C}{iv}-\ion{Mg}{ii}) and FWHM(\ion{C}{iv}/\ion{Mg}{ii}) relation is illustrated in \cref{fig:civmgiidiffv+ratiofwhm+hbfwhm_t12s1216}. The $\Delta v$(\ion{C}{iv}-\ion{Mg}{ii}) and FWHM H$\beta$ are positively correlated, while the opposite correlation between FWHM(\ion{C}{iv}/\ion{Mg}{ii}) and FWHM H$\beta$ is seen (\cref{tab:spearman}). This finding agrees with the expectation from the orientation indicator using H$\beta$ line width, whereby the FWHM of H$\beta$ is broader for high inclination, less blueshifted, and small FWHM(\ion{C}{iv}/\ion{Mg}{ii}) ratio objects.

Using T12 sample, the $\Delta v$(\ion{C}{iv}-\ion{Mg}{ii}) and FWHM(\ion{C}{iv}/\ion{Mg}{ii}) correlation is weaker compared to that using S12xS16 sample, though both are significant at $p_{S}<0.05$. The anti-correlation between the line width ratio and H$\beta$ is highly significant with $p_{S} \ll 0.001$ in both samples. Meanwhile, the relationship between the velocity shift and H$\beta$ is not statistically significant in T12 sample but significant at $p_{S}<0.05$ in S12xS16 sample. Combining both samples yields a significant $p_{S}$ value between the parameters.

\subsection{Line Strength of \texorpdfstring{[\ion{O}{iii}]}{[OIII]}}

\begin{figure}
\centering
\includegraphics[width=0.48\textwidth,keepaspectratio]{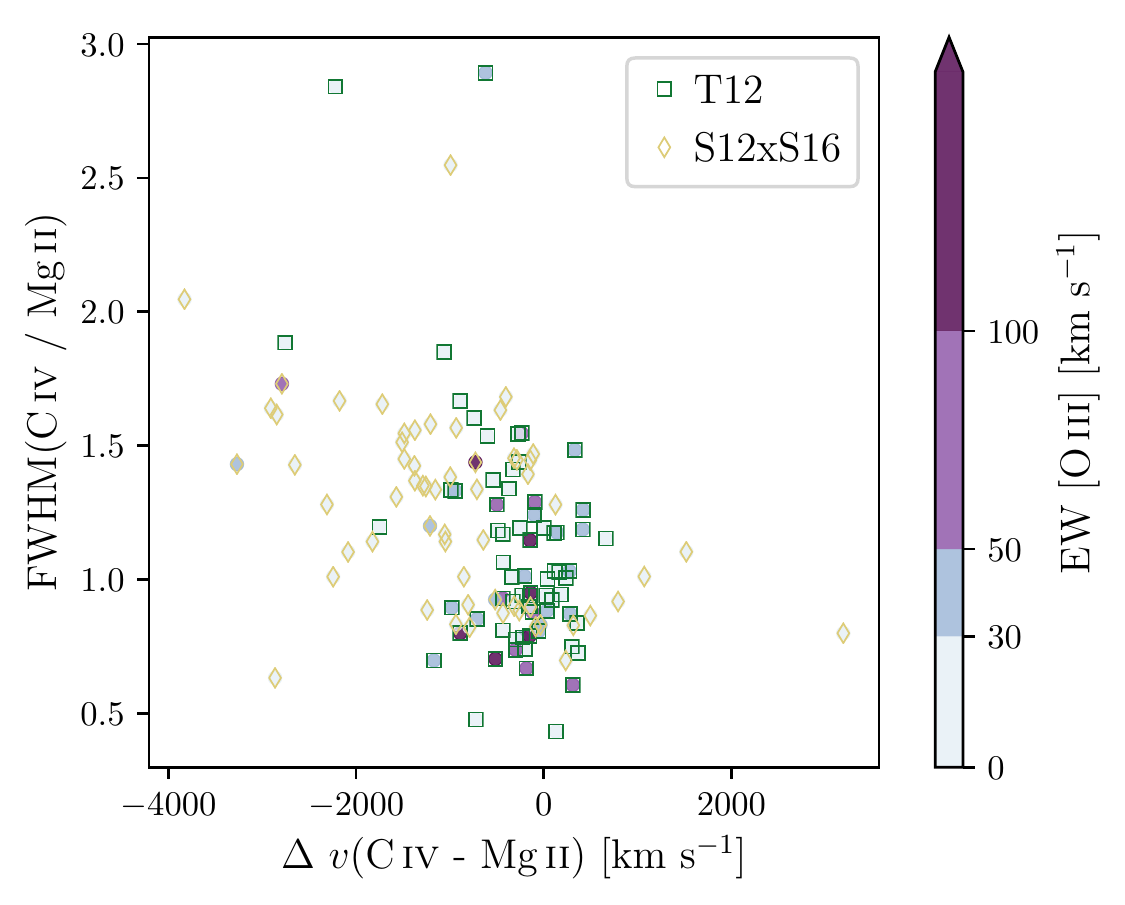}
\caption{Ratio of full width at half maximum, FWHM, against velocity shift, $\Delta v$, between \ion{C}{iv} and \ion{Mg}{ii}, with equivalent width (EW) [\ion{O}{iii}]. The distribution is colour-coded by increasing colour gradient with EW [\ion{O}{iii}] using quasars sample of T12 \citep{Tang+:2012} in green squares and S12xS16 \citep{Shen+Liu:2012,Shen:2016} in yellow diamonds.}
\label{fig:civmgiidiffv+ratiofwhm+oiiiew_t12s1216}
\end{figure}

We use the bright quasar samples \citep{Tang+:2012} and the SDSS DR7Q optical and near-infrared spectral measurements \citep{Shen+Liu:2012,Shen:2016} described in the previous section. The behaviour of EW [\ion{O}{iii}] on the FWHM(\ion{C}{iv}/\ion{Mg}{ii}) versus $\Delta v$(\ion{C}{iv}-\ion{Mg}{ii}) map, shown in \cref{fig:civmgiidiffv+ratiofwhm+oiiiew_t12s1216}, seems to be consistent with high EW [\ion{O}{iii}] objects situated at lower right end of the map for viewing angle close to the accretion disk. When considering both samples from T12 and S12xS16, the Spearman rank correlation coefficients of velocity shifts and FWHM ratio with EW [\ion{O}{iii}] are highly significant ($p_{S}<0.001$) with $r_{S}=0.304$ and $r_{S}=-0.399$, respectively (\cref{tab:spearman}).

However, examining the samples separately, the EW [\ion{O}{iii}] with $\Delta v$(\ion{C}{iv}-\ion{Mg}{ii}) and FWHM(\ion{C}{iv}/\ion{Mg}{ii}) relations in S12xS16 sample are not significant. This inconsistency is probably due to the lack of high EW [\ion{O}{iii}] quasars in S12xS16 sample to be considered as near edge-on orientation objects, whereby only 6 quasars have EW [\ion{O}{iii}] $\geq 30\,$\AA, with 4 of them range $\sim 30 \text{--} 40\,$\AA. This is portrayed in \cref{fig:civmgiidiffv+ratiofwhm+oiiiew_t12s1216}, where it is colour-coded into a different range of EW for distinction. The sample in T12 contains more objects with high EW, with almost half or 30 of them that display high EW [\ion{O}{iii}] of $\geq 30\,$\AA. There is a highly significant anti-correlation between the EW [\ion{O}{iii}] and FWHM(\ion{C}{iv}/\ion{Mg}{ii}) at $p_{S}<0.001$, though EW [\ion{O}{iii}] and $\Delta v$(\ion{C}{iv}-\ion{Mg}{ii}) show non-significant correlation. This might be also because of the weak negative relation between $\Delta v$(\ion{C}{iv}-\ion{Mg}{ii}) and FWHM(\ion{C}{iv}/\ion{Mg}{ii}) with $r_{S}=-0.271$ and lower significance at $p_{S}<0.05$.

Assuming that the line width of H$\beta$ and the line strength of [\ion{O}{iii}] are both inclination diagnostics, the Spearman correlation coefficients are highly significant at $p_{S} \ll 0.001$ when using T12 sample and with S12xS16 merged. The $r_{S}$ and $p_{S}$ values are weaker in S12xS16 sample.

\subsection{Low-ionisation Broad Absorption Line Quasars}

To examine the emission line properties of LoBAL quasars, we use the catalogue of BAL quasars from the SDSS DR3 \citep{Trump+:2006}. The descriptions on the construction of the spectra and selection of BAL quasars are given in Sect.~3 and Sect.~4 of their paper. They identified $0.5 \leq z \leq 2.15$ quasars with LoBAL features in \ion{Mg}{ii} line that satisfy a less stricter measure of BAL, the absorption index \citep[AI;][]{Hall+:2002}. They adopted a modified version of AI such that it measures the true EW and the whole absorption up to 29\,000\,\kms\ through an automated pipeline. Their parent sample consists of about 1.31 per cent LoBAL quasars. By cross-matching their catalogue with the quasar properties SDSS DR7Q catalogue \citep{Shen+:2011}, we obtain 72 LoBAL quasars.

\begin{figure}
\centering
\includegraphics[width=0.48\textwidth,keepaspectratio]{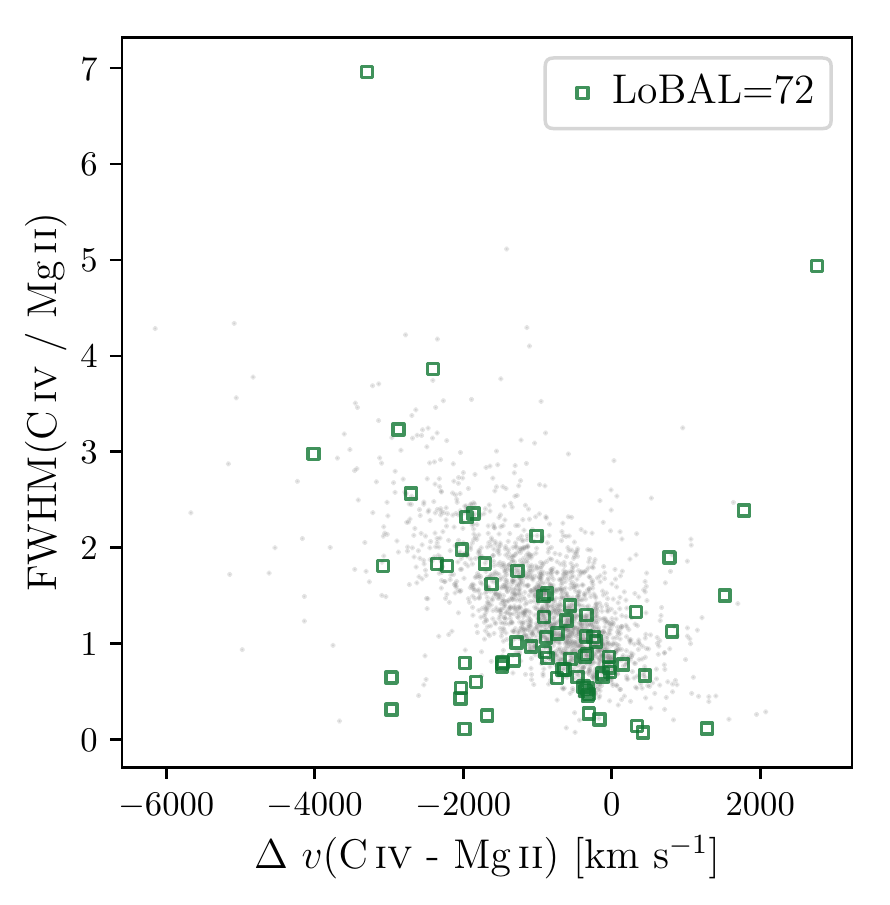}
\caption{Ratio of full width at half maximum, FWHM, against velocity shift, $\Delta v$, between \ion{C}{iv} and \ion{Mg}{ii}, with low-ionisation broad absorption line (LoBAL) quasars. The LoBAL classification is obtained from \citet{Trump+:2006}. The emission lines measurements are obtained from \citet{Shen+:2011} SDSS DR7Q catalogue. For comparison, their high spectral quality (high S/N) samples are superimposed in small grey symbols.}
\label{fig:civmgiidiffv+ratiofwhm+lobal_t06xs11}
\end{figure}

\Cref{fig:civmgiidiffv+ratiofwhm+lobal_t06xs11} shows the sample of LoBAL quasars on the ratio FWHM and velocity shift relation. They seem to be preferentially clustered towards the lower right of the distribution, implying mid to high inclination angles. This supports our viewpoint that LoBALs are seen preferentially near equatorial angle, though about 20 per cent appears on the opposite end with more blueshifted and higher \ion{C}{iv} FWHM than \ion{Mg}{ii}. It is crucial to point out that BALs might inherently have large blueshift, which reflects the wind outflow. Additionally, the absorption trough might affect the line width measurement.

\subsection{Summary of Comparison with Other Inclination Measurements}

The presented statistics and analyses demonstrate that our proposed correlation using the ratio of FWHM and the velocity shift of \ion{C}{iv} and \ion{Mg}{ii} is generally consistent with other inclination angle measurements from the literature. Inclinations of radio sources can be loosely inferred from the radio morphology and radio core dominance. These measurements of inclination are also roughly consistent when projected onto our proposed indicator.

Since H$\beta$ and \ion{Mg}{ii} are both low-ionisation lines, they are expected to be emitted at similar locations in the BLR, close to the disk with a dominant rotational component. Hence the strong correlation between the FWHM(H$\beta$) and ratio FWHM(\ion{C}{iv}/\ion{Mg}{ii}). The strength of [\ion{O}{iii}] provides an indication of the geometry of the emitting region, which also reflects orientation. This metric is weakly correlated when compared to the FWHM ratio and velocity shifts. The distribution of LoBAL quasars on our proposed inclination indicator also agrees with the prediction that these population being observed close to equatorial plane.

\section{Comparison with Simulations} \label{sec:compsim}

Since our proposed inclination indicator is derived mainly based on the kinematics of the wind, we can test whether our simple kinematical BLR disk-wind model in \citet{Yong+:2016,Yong+:2017} is able to predict a similar trend as seen in observation. Detailed descriptions on the wind kinematics and modelling are mentioned in \citet{Yong+:2016,Yong+:2017}, and will be briefly included here.

\subsection{Modelling the Wind Kinematics}

\begin{figure*}
\centering
\includegraphics{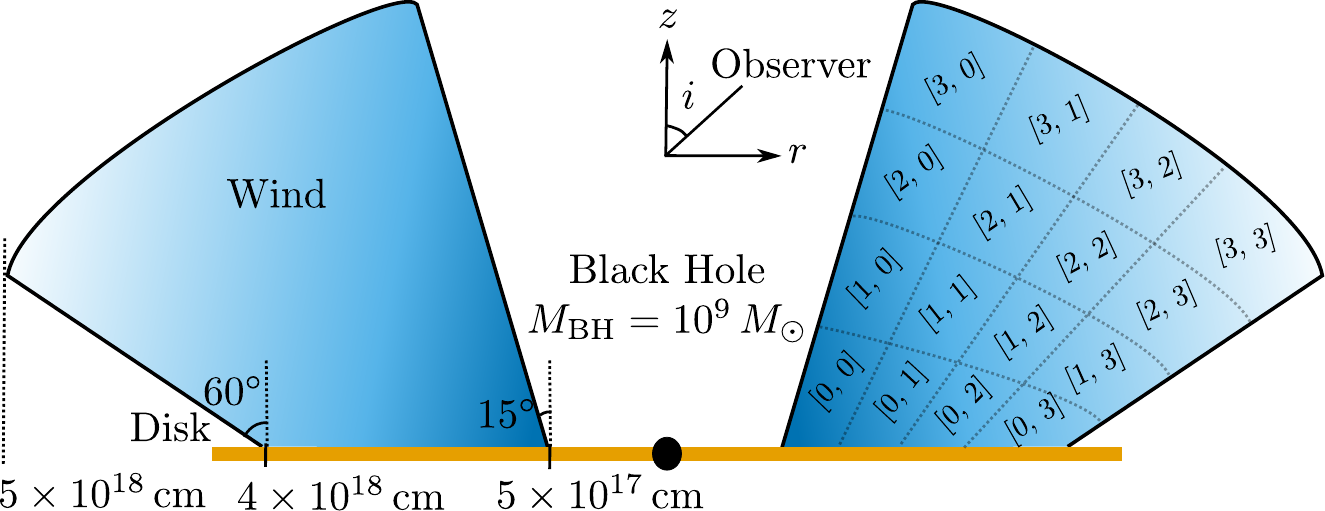}
\caption{A sketch of the disk-wind model with wide opening wind. {\em Left}: Adopted fiducial parameters values. {\em Right}: Wind zones indicated by rows and column, from bottom to top and left to right.}
\label{fig:cywidewind_paramzone}
\end{figure*}

The kinematics of the wind is adopted from \citet{Shlosman+Vitello:1993} with radiative transfer in Sobolev limit \citep{Rybicki+Hummer:1978,Rybicki+Hummer:1983}. A sketch of the model is presented in \cref{fig:cywidewind_paramzone}. The geometry of the disk-wind BLR model is assumed to be axially symmetric and is expressed in cylindrical coordinates $(r,\phi,z)$. The radial and azimuthal component on the accretion disk plane are $r$ and $\phi$. The disk rotation axis is defined at $z$ and the angle it forms with the line-of-sight is the inclination angle, $i$.

The accretion disk is flat and geometrically thin, but optically thick. The outflowing wind arises from the disk and spirals outward in a helical trajectory. The total velocity of the wind consists of the rotational (on accretion disk plane) and poloidal (on $rz$-plane) component. Close to the ionising source, the wind velocity is dominated by rotational component. As the poloidal distance increases, the poloidal velocity component gradually amplifies.

The main parameter values adopted are shown on the left side of \cref{fig:cywidewind_paramzone}. To create our fiducial model, we utilise constraints from observations and theory. The black hole mass is estimated based on the virial black hole mass measurements from the SDSS DR7Q quasar properties catalogue \citep{Shen+:2011}. Assuming that it scales with the continuum luminosity, $\lambda L_{\lambda}$, and the broad emission line width, the virial black hole mass, \subtext{M}{BH,vir}, is expressed as
\begin{align}
\log \subtext{M}{BH,vir} = a\log\left(\frac{\lambda L_{\lambda}}{10^{44}\,\ergs}\right) + 2\log\mathrm{FWHM} + b,
\end{align}
where $(a,b)=(0.62,0.74)$ calibrated based on \ion{Mg}{ii} line for $0.7 \leq z < 1.9$ \citep{Shen+:2011} and $(a,b)=(0.53,0.66)$ calibrated based on \ion{C}{iv} line for $z \geq 1.9$ \citep{Vestergaard+Peterson:2006}. The mean of \subtext{M}{BH,vir} in our SDSS DR7Q sample is $10^{9.49}\,M_{\odot}$ ($3.07 \times 10^{9}\,M_{\odot}$), and we set $\subtext{M}{BH}=10^{9}\,M_{\odot}$ in our model.

The mass accretion rate onto the black hole is determined by the accretion parameters. The radiative efficiency at which mass is converted to radiation is given by $\eta=\subtext{L}{bol}/\subtext{\dot{M}}{acc} c^{2}$, where \subtext{L}{bol} is the bolometric luminosity and \subtext{\dot{M}}{acc} is the total mass accretion rate. For luminous sources at intermediate redshift, $\eta$ is found to be $>0.2$ \citep{Davis+Laor:2011,Trakhtenbrot:2014}. Since our sample consists of high bolometric luminosity quasars with mean $\subtext{L}{bol} \approx 1.23 \times 10^{47}\,\ergs$, we use $\eta=0.2$. The accretion rate is then $\subtext{\dot{M}}{acc} \sim 10\,M_{\odot}\,$yr$^{-1}$ and the total mass-loss rate of the wind, \subtext{\dot{M}}{wind}, is also set to this value.

Following the disk-wind model proposed in \citet{Yong+:2018}, the wind has a wide range of angles from $15\degree$ to $60\degree$. The wind is also optically thick and is bounded within radius from $5 \times 10^{17}\,$cm ($\approx 3385\,\subtext{r}{g}$) to $4 \times 10^{18}\,$cm ($\approx 27\,082\,\subtext{r}{g}$), where $\subtext{r}{g}=G\subtext{M}{BH}/c^{2}$ is the gravitational radius, $G$ is the gravitational constant, and $c$ is the speed of light. To justify our selected parameters, these values are chosen using the well-known BLR radius--luminosity ($r \text{--} L$) relation \citep{Kaspi+:2007,Bentz+:2009} from reverberation mapping studies as a baseline. The BLR radius in units of light days is given by
\begin{align}
\subtext{r}{BLR}=c\left(\frac{\lambda L_{\lambda}}{10^{44}\,\ergs}\right)^{d},
\end{align}
where $(c,d)=(19.1,0.56)$ derived at 1350\,\AA\ using the average of fitted slopes from mean time lags of Balmer lines for all dataset per object \citep{Kaspi+:2005} and $(c,d)=(18.5,0.62)$ derived at 3000\,\AA\ \citep{McLure+Dunlop:2004}. Using these relations, the mean radii of the BLR are $1.39 \times 10^{18}\,$cm for 1350\,\AA\ and $1.50 \times 10^{18}\,$cm for 3000\,\AA, which provide an estimate of the emitting region for \ion{C}{iv} and \ion{Mg}{ii} lines in the wind. These values are within the bounds of the BLR wind in our model. The size of BLR using H$\beta$ of reverberation-mapped samples is also found to be $\sim 2\text{--}3$ times larger than that using \ion{C}{iv} \citep{Peterson+:2004}. Subsequently, we fix the BLR radius to be $\subtext{r}{BLR}=5 \times 10^{18}\,$cm ($\approx 33\,852\,\subtext{r}{g}$).

Studies have found that the properties of emission lines differ depending on the ionisation state of the line. The majority of high-ionisation lines, such as \ion{C}{iv}, exhibit broader line width \citep{Osterbrock+Shuder:1982,Mathews+Wampler:1985} and larger blueshift \citep[e.g.,][]{Gaskell:1982} than the low-ionisation lines, such as \ion{Mg}{ii}. This suggests that the structure of BLR is stratified with different ionisation lines emitting from different regions in the BLR \citep{Peterson+Wandel:1999,Kollatschny:2003,Peterson+:2004}. Low-ionisation lines are closer to the accretion disk base but far from the centre of ionising source, where the rotational velocity dominates. In comparison, high-ionisation lines are radially nearer to the central ionising source but further away in poloidal distance with more contribution from the poloidal velocity.

For \ion{C}{iv} to be blueshifted and broader than \ion{Mg}{ii}, the wind poloidal velocity has to be boosted enough to overcome the rotational component. There are couple of free parameters that regulate the poloidal velocity, two of which are $R_{v}$ and $\alpha$. The acceleration scale height, $R_{v}$, of the poloidal velocity determines the stage where the wind attains half of its terminal velocity. The acceleration of the poloidal wind is dictated by the power law index, $\alpha$. The quantities $R_{v}$ and $\alpha$ are set to be $1 \times 10^{18}\,$cm ($\approx 6770\,\subtext{r}{g}$) and 3.5 respectively such that the poloidal velocity near the wind base accelerates slowly but quickly gains speed with higher poloidal distance.

To emulate the stratification in the BLR wind, the wind is partitioned into 4 by 4 `wind zones', labelled based on the position of its row and column [$a,b$], as illustrated on the right side of \cref{fig:cywidewind_paramzone}. The assigned numbers begin from the zone nearest to the black hole with increasing poloidal distance. We explore a set of zone pairs for \ion{Mg}{ii} and \ion{C}{iv} emission lines. Wind zones [0, 1]--[0, 3] are chosen as plausible line emitting region for \ion{Mg}{ii} line, and zones [2, 0]--[2, 1] for \ion{C}{iv}.

Once the wind kinematics are set up, a Monte Carlo simulation is employed to generate large number of random particles within each zone. The projected line-of-sight velocity as a function of inclination angle between $10\degree$ and $80\degree$ is computed for each particle, which is binned to create a histogram, and finally considered in the production of the total line profile. Within each specified bins, the counts are weighted according to the density and radiative transfer intensity. The contribution of every particle at its given velocity is then the total emission line profile. A smooth emission line profile is produced by convolving with a Gaussian kernel. The line width of the smoothed line is evaluated at its FWHM. The velocity shift of individual emission line is estimated from the median of the line to the axis centre. The median is used instead of the line peak due to the shape of the generated line profile. The velocity shift of \ion{C}{iv} with respect to \ion{Mg}{ii} is then the difference between the two shifts.

\subsection{Predictions with Simulation}

\begin{figure*}
\centering
  \begin{subfigure}[t]{0.495\textwidth}
  \includegraphics[width=\textwidth]{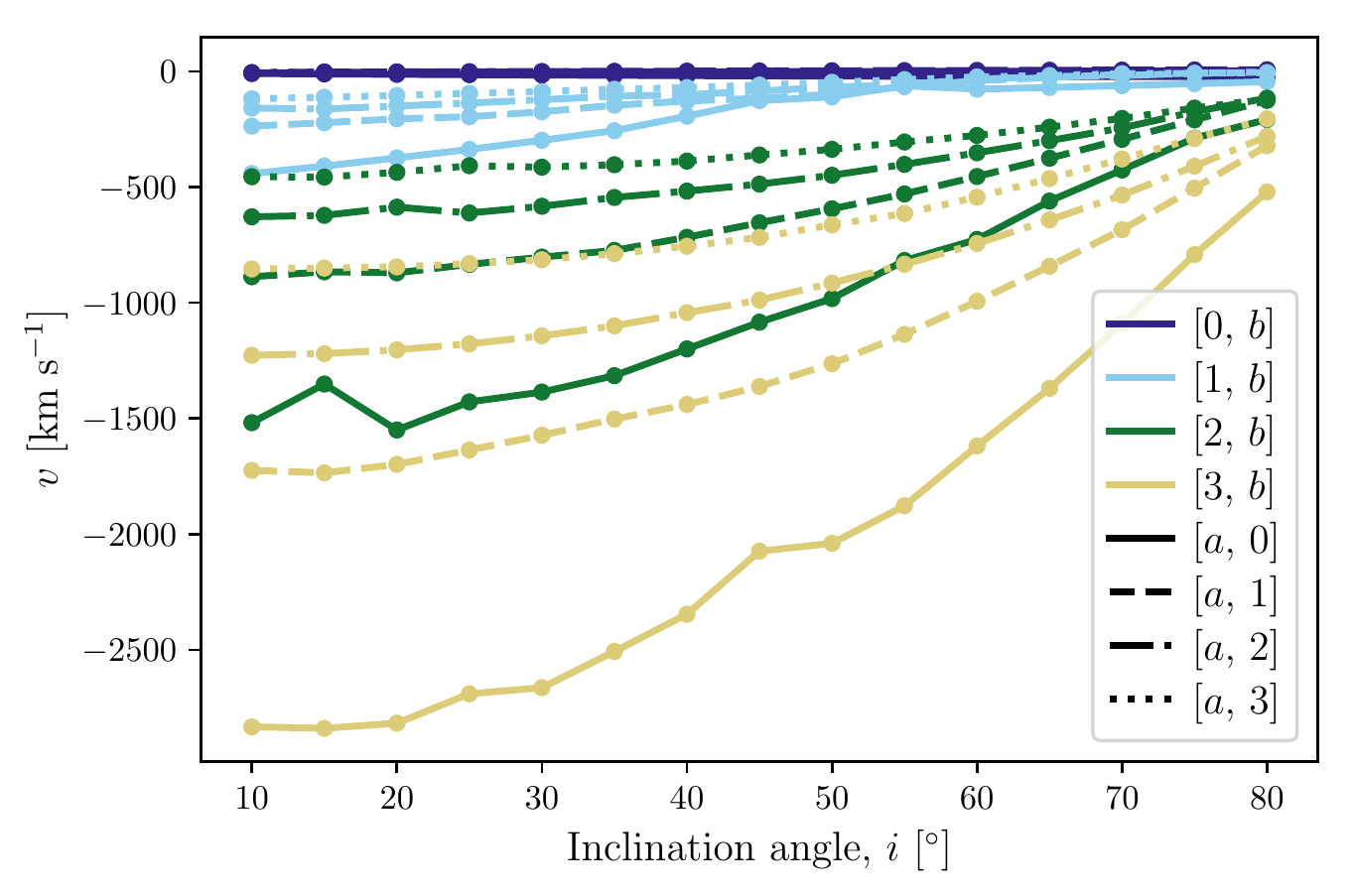}
  \caption{Velocity shift relative to the line profile centroid, $v$, against $i$}
  \label{fig:iangle+vshift_t15-60_ab}
  \end{subfigure}%
  \begin{subfigure}[t]{0.495\textwidth}
  \includegraphics[width=\textwidth]{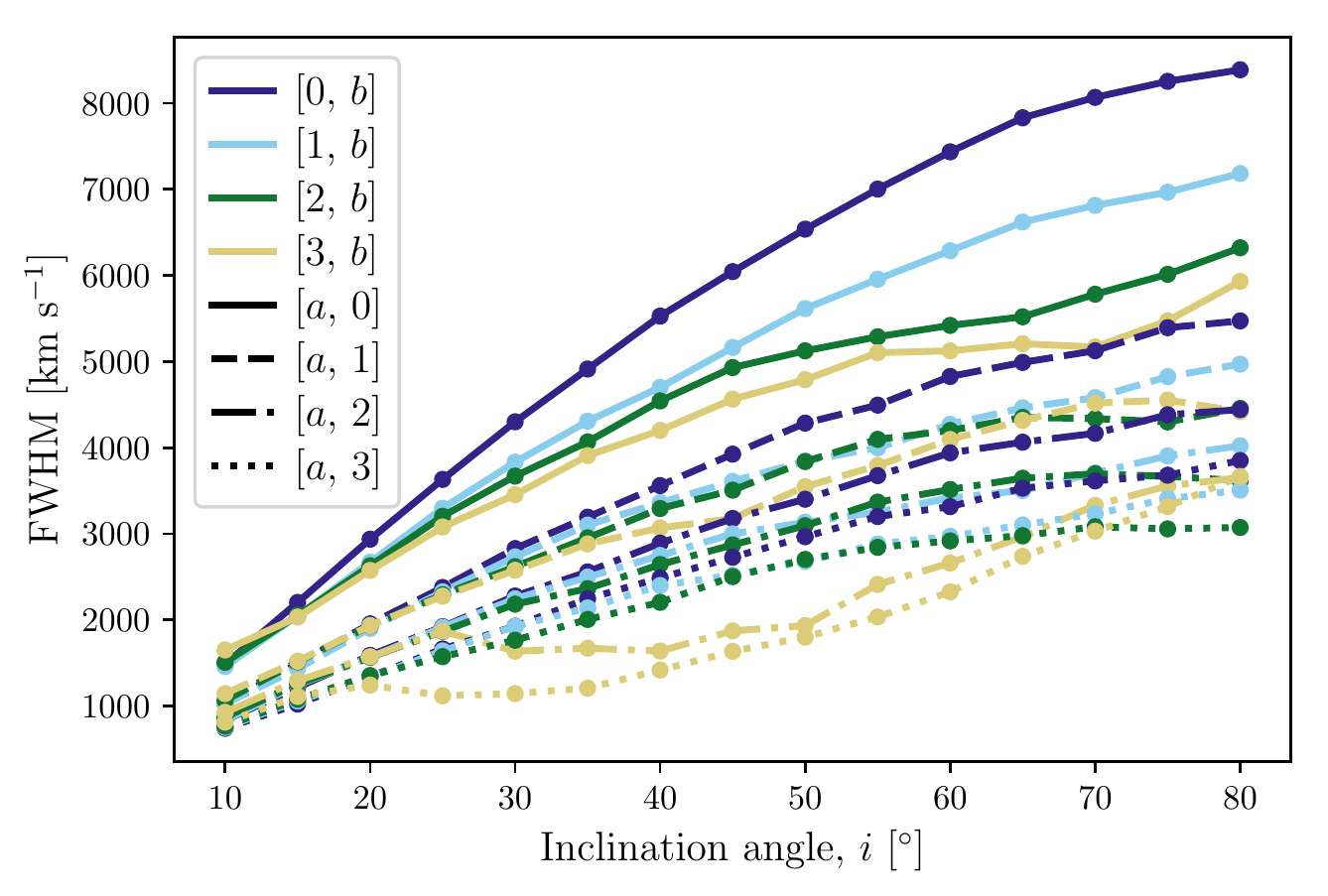}
  \caption{Full width at half maximum, FWHM, against $i$}
  \label{fig:iangle+fwhm_t15-60_ab}
  \end{subfigure}
\caption{Simulated emission line properties as a function of inclination angle $i$. The different colours and line-styles represent distinct rows and columns in the wind zones.}
\label{fig:iangle+_t15-60_ab}
\end{figure*}

The predicted emission line profile properties with inclination angle are shown in \cref{fig:iangle+_t15-60_ab}. As expected, the line profiles closest to the base of the wind streamline at zones [0, $b$] are less blueshifted and broader due to the stronger contribution from the Keplerian rotational velocity. Meanwhile, emission lines situated at zones further above the disk have larger wind poloidal velocity wind and lesser from the rotational component, which are reflected by the more blueward shifts away from the line centroid and reduced FWHMs. The line shifts and FWHMs in zones [2, $b$] and [3, $b$] are also fairly similar.

The blueshift is the most prominent at low inclination angle (\cref{fig:iangle+vshift_t15-60_ab}). Subsequently, it becomes smaller with increasing viewing angle since the line-of-sight is moving away relative to the wind. At small inclination angle, the line-of-sight velocity is also less affected by the rotational component, and hence the smaller FWHM (\cref{fig:iangle+fwhm_t15-60_ab}). The line width peaks as the angle of inclination approaches equatorial view.

\begin{figure}
\centering
\includegraphics[width=0.48\textwidth,keepaspectratio]{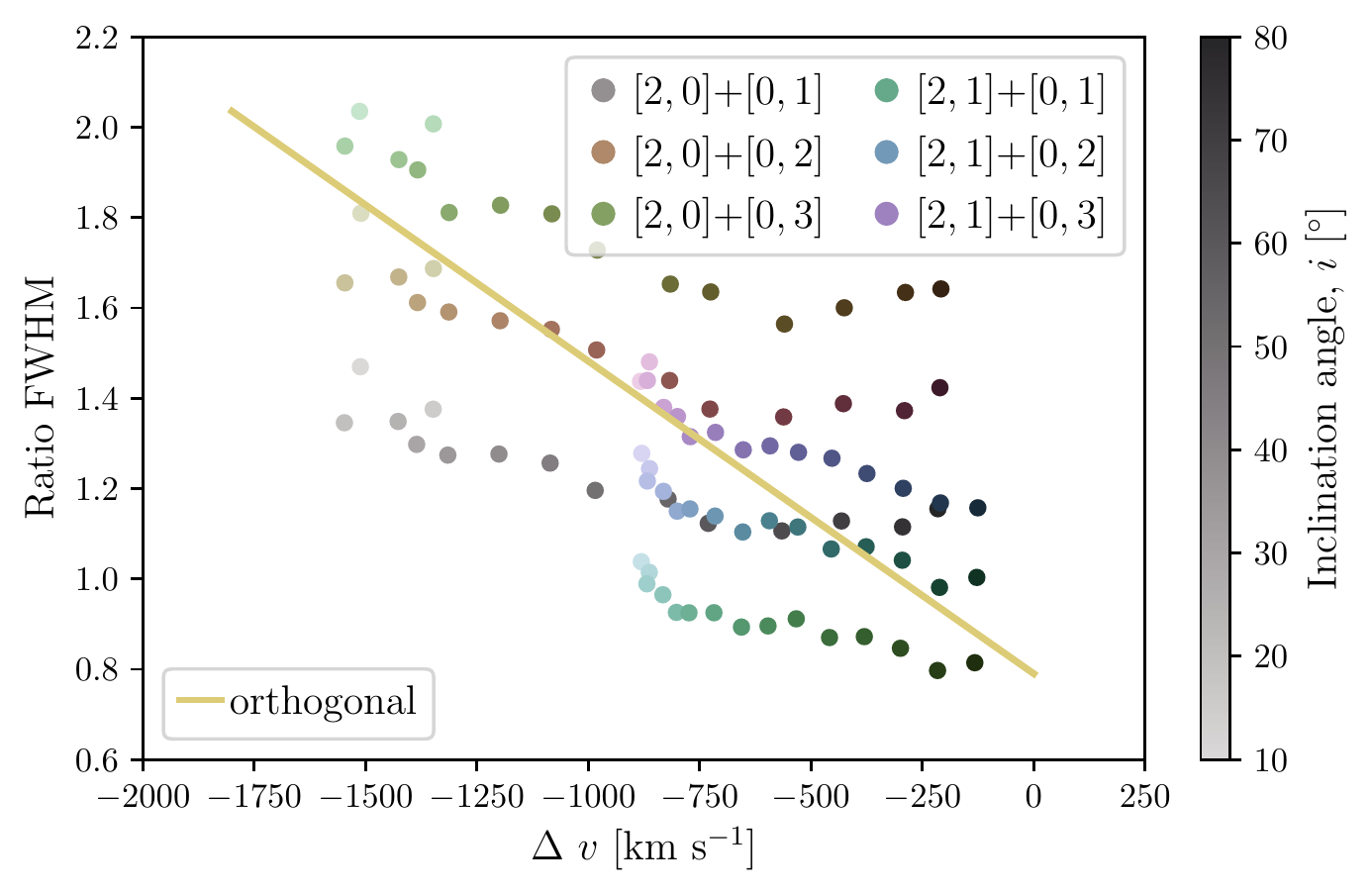}
\caption{Simulated ratio of full width at half maximum, FWHM, against velocity shift, $\Delta v$, between two emission lines in distinct wind zones. The different pairs of wind zones [2, 0]--[2, 1] relative to [0, 1]--[0, 3] are represented by the different colours. The increasing colour gradient indicates the inclination, $i$, from low at $10\degree$ (near pole-on) to high at $80\degree$ (near edge-on). For comparison, the fitted line from observation is superimposed in coloured lines.}
\label{fig:diffvshift+ratiofwhm_t15-60_bcesort_dr7q}
\end{figure}

\Cref{fig:diffvshift+ratiofwhm_t15-60_bcesort_dr7q} presents the ratio FWHM versus the velocity shifts of the line profiles emitted from zones [2, 0]--[2, 1] and [0, 1]--[0, 3]. As explained in the previous section, [2, 0]--[2, 1] is the zone of the wind emitting \ion{C}{iv}, while [0, 1]--[0, 3] is emitting \ion{Mg}{ii}. \Cref{fig:diffvshift+ratiofwhm_t15-60_bcesort_dr7q} therefore examine how the ratio of the FWHM and the velocity shifts between the two lines vary when comparing different regions of the wind. As the relative distance between the two emitting regions increases, the velocity shift decreases while the FWHM ratio increases. This is demonstrated in, for example, the line profiles in zone pairs [2, 0] and [0, 2] have higher blueshift and FWHM ratio than those in zones [2, 1] and [0, 2].

As a comparison, the fitted linear regression model using observational data is also overlaid in \cref{fig:diffvshift+ratiofwhm_t15-60_bcesort_dr7q}. The prediction from our simple kinematical model matches well with the fitted slopes, with the scales of the parameter within those from observations. In general, the model is able to qualitatively reproduce the negative relationship between the ratio FWHM and velocity shift with higher inclination angle (near edge-on), consistent with our proposition.

\section{Potential of the Orientation Indicator} \label{sec:robusti}

Using two easily measured characteristics of emission line in the UV-optical, we demonstrate that the correlation between the velocity difference and FWHM ratio of \ion{C}{iv} and \ion{Mg}{ii} lines is due to angle-of-viewing. We argued that the relation can be explained in a simple physical model of the disk-wind. We have tested the predictions of our diagnostic against other tests which claim to indicate orientation. The results shown indicate broad consistency with our indicator. The analytic disk-wind modelling performed has also successfully reproduced the range of velocity difference and FWHM ratio.

All evidence that has been presented favours a disk-wind model similar to the one presented. However, as mentioned in \cref{ssec:errfits}, there are a few caveats in our approach that are worth reiterating. The dataset used to test the model has been derived from automated algorithms of the very large SDSS dataset. Thus, it is expected that some individual data points will be revised with more careful individual analysis. However, this will not void the basic trend. The chosen linearity between the projected distance and the inclination angle and the range of this mapping is also not entirely justified, however, this choice leads to a physically sensible interpretation and fits our simple model. Ultimately, further modelling, including the choice of the \ion{C}{iv} and \ion{Mg}{ii} emission regions and photoionisation effects, will be needed to better establish this mapping.

The expected variation in physical properties due to the angle-of-viewing provides a strong motivation to develop this model and improve the estimation. Future detailed analysis of the datasets, combined with optimisation of model parameters and properly account for other intrinsic properties of the source such as the accretion state and luminosity, will enable the angle-of-viewing to be well-determined. Once the effects of orientation are accounted for, more robust determinations of other physical parameters are possible, providing a natural framework for unravelling a detailed physical model for quasars. If the physics of these highly luminous sources can be unravelled, they potentially provide a standard candle, with which to map the structure of the distant universe \citep{Marziani+Sulentic:2014,Risaliti+Lusso:2015,Risaliti+Lusso:2019,Lusso+:2019}.

The measured characteristics of the high and low ionisation lines differ, providing clear evidence that they arise in different geometric and kinematic locations of the disk-wind. Interpreted in the correct model, this data can be used to measure angle-of-viewing to the quasar. If the angle-of-viewing can be robustly measured, then finally a key observational parameter in the physical model of the emitting regions of a quasar can be determined, significantly helping the mapping, both geometrically and kinematically, of those regions. Also, emission line widths are used to estimate black hole masses, and the orientation may introduce up to an order of magnitude variation \citep{Yong+:2016}. Thus, determination of the angle-of-viewing will significantly improve black hole estimates.

\section{Summary} \label{sec:summary}

The scales of AGN emission regions are small enough that the detailed structure is unlikely to be directly resolved at wavelengths shorter than infrared in the foreseeable future. Thus, the challenge of understanding their complex physical structure must rely on observational proxies and physically motivated modelling. Since AGN are powered by gravitational accretion, the emission regions are expected to be axisymmetric. This also suggests that the accretion disk will be opaque in the inner regions, and therefore only the forward side of the emission region will be observed. Therefore, unravelling the detailed physics will be greatly simplified if an accurate measure of the viewing angle of the observer can be determined.

In this work, we have made significant progress in determining the viewing angle to individual AGN. We have:
\begin{itemize}[leftmargin=.125in,itemsep=1ex]
  \item Found a strong correlation between two easily measurable parameters in the ultraviolet emission spectrum of AGN. Specifically, we have chosen a high ionisation line, \ion{C}{iv}, and a lower ionisation line, \ion{Mg}{ii}, and shown that they must be emitted in different parts of a disk-wind. The measured physical parameters are the velocity shifts and line widths, both of which might be expected to depend on orientation.
  \item Established that qualitatively, the correlation is consistent with an explanation due to orientation angle. Quasars viewed at close to face-on angle are predicted to exhibit large blueshifts and line width ratio of \ion{C}{iv} and \ion{Mg}{ii}. In contrast, the blueshifts and relative ratio of the line widths decrease as the inclination is towards edge-on.
  \item Compared this measurement with a few known orientation indicators in the literature. In particular, the results for LoBALs is strongly consistent, but other indicators are broadly consistent with our predictions.
  \item Modelled the correlation using a simple disk-wind, suggesting that a detailed exploration of the model parameters will further refine the structure of the physical models for the broad emission line regions of AGN.
\end{itemize}

AGN, particularly at high redshift, have a profound impact on galaxy evolution and as tracers of cosmography. If a standard physical model for the observables of an AGN can be determined, then these objects might realise their potential for cosmological studies.

\section*{Acknowledgements}

We thank the anonymous referee for valuable suggestions on the manuscript. In addition, we thank Tiziana Di Matteo and Tamara Davis for their helpful comments. This research has made use of the VizieR catalog access tool, CDS, Strasbourg, France. The original description of the VizieR service was published in \citep{Ochsenbein+:2000}. This research made use of the Python libraries including open source packages such as \texttt{astropy} \citep{Astropy:2013}, \texttt{matplotlib} \citep{Hunter:2007}, \texttt{numpy} \citep{vanderWalt+:2011}, and \texttt{scipy} \citep{Jones+:2001}.


\bsp	
\label{lastpage}
\end{document}